\documentclass[a4paper,UKenglish,cleveref, autoref, thm-restate]{lipics-v2021}

\usepackage{caption}
\usepackage{subcaption}
\usepackage{todonotes}
\usepackage{multicol}
\usepackage{amssymb}
\usepackage{bussproofs}
\usepackage{mathtools}
\usepackage{mathpartir}
\usepackage{amsthm}

\usepackage{tikz-cd}
\usepackage{tikz}
\usetikzlibrary{positioning}

\usetikzlibrary{backgrounds,fit,decorations.pathreplacing}  
\newcommand{\rbox}[1]{\raisebox{-.8ex}{\scalebox{.7}{\begin{tikzpicture}\draw (0,0) circle [radius=0.3] node {$#1$};\end{tikzpicture}}}}

\usetikzlibrary{arrows,positioning} 
\tikzset{
    >=stealth',
    punkt/.style={
           rectangle,
           rounded corners,
           draw=black, very thick,
           text width=6.5em,
           minimum height=2em,
           text centered},
    pil/.style={
           ->,
           thick,
           shorten <=2pt,
           shorten >=2pt,}
}
\tikzstyle{operator} = [draw,fill=white,minimum size=1.5em]
\tikzstyle{measure} = [draw,fill=white,minimum size=1.5em,path picture = 
{\draw[black] ([shift={(.1,.25)}]path picture bounding box.south west) to[bend left=50] ([shift={(-.1,.25)}]path picture bounding box.south east);
\draw[black,-latex] ([shift={(0,.1)}]path picture bounding box.south) -- ([shift={(.2,-.1)}]path picture bounding box.north);}]
\tikzstyle{surround} = [fill=green!10,thick,draw=black,rounded corners=2mm]

\newcommand{\denot}[1]{\llbracket #1 \rrbracket}  
\newcommand{\bM}[0]{\overline{M}}                 
\newcommand{\bbM}[0]{\overline{\overline{M}}}     
\newcommand{\bbMF}[0]{\overline{\overline{M}}_F}  

\newcommand{\llbracket}{[\![}
\newcommand{\rrbracket}{]\!]}

\title{Concrete Categorical Model of a Quantum Circuit Description Language with Measurement}

\titlerunning{Categorical Model for a Quantum Language with Measurement}

\author{Dongho Lee}{Université Paris-Saclay, CentraleSupélec, LMF, France \& CEA, List, France \and \url{https://lmf.cnrs.fr/Perso/Dongho_Lee} }{fredldh@gmail.com}{}{}

\author{Valentin Perrelle}{Université Paris-Saclay, CEA, List, France }{Valentin.PERRELLE@cea.fr}{}{}

\author{Benoît Valiron}{Université Paris-Saclay, CentraleSupélec, LMF,
  France \and \url{https://www.monoidal.net} }{benoit.valiron@lri.fr}{}{}

\author{Zhaowei Xu}{Université Paris-Saclay, LMF, France}{zhaowei@lri.fr}{}{}

\authorrunning{D. Lee, V. Perrelle, B. Valiron, Z. Xu}

\Copyright{Dongho Lee, Valentin Perrelle, Benoît Valiron and Zhaowei Xu}



\category{} 

\relatedversion{} 

\acknowledgements{This work was supported in part by the French
  National Research Agency (ANR) under the research projects SoftQPRO
  ANR-17-CE25-0009-02 and PPS ANR-19-CE48-0014, by the DGE of the
  French Ministry of Industry under the research project
  PIA-GDN/QuantEx P163746-484124, and by STICAMSUD 21-SITC-20
  Qapla'. The authors want to thank Christophe Chareton
  and Sébastien Bardin for enlightening discussions.}

\EventEditors{Miko{\l}aj Boja\'{n}czyk and Chandra Chekuri}
\EventNoEds{2}
\EventLongTitle{41st IARCS Annual Conference on Foundations of Software Technology and Theoretical Computer Science (FSTTCS 2021)}
\EventShortTitle{FSTTCS 2021}
\EventAcronym{FSTTCS}
\EventYear{2021}
\EventDate{December 15--17, 2021}
\EventLocation{Virtual Conference}
\EventLogo{}
\SeriesVolume{213}
\ArticleNo{13}

\begin{document}

\nolinenumbers

\maketitle

\begin{abstract}
  In this paper, we introduce dynamic lifting to a quantum
  circuit-description language, following the Proto-Quipper language
  approach. Dynamic lifting allows programs to transfer the result of
  measuring quantum data ---qubits--- into classical data ---booleans---.
  We propose a type system and an operational semantics for the
  language and we state safety properties. Next, we introduce a concrete
  categorical semantics for the proposed language, basing our approach
  on a recent model from Rios\&Selinger for Proto-Quipper-M. Our
  approach is to construct on top of a concrete category of circuits
  with measurements a Kleisli category, capturing as a side effect the
  action of retrieving classical content out of a quantum memory. We
  then show a  soundness result for this semantics.
\end{abstract}

\section{Introduction}
\label{s:introduction}

In quantum computation, one considers a special kind of memory where
data is encoded on the state of objects governed by the laws of
quantum mechanics. The basic unit for quantum data is the quantum bit,
or qubit, and in general, a quantum memory is understood as consisting
in individually addressable qubits. As derived in the no-cloning
theorem~\cite{wootters82single}, qubits are
\emph{non-duplicable} objects.
The state of a quantum memory can
be represented by a unit vector in a complex Hilbert space.
Elementary operations on qubits consist in unitary operations on the
state space, called \emph{quantum gates}, and \emph{measurements},
which are probabilistic operations returning a classical boolean.

The usual model for quantum computation is the notion of quantum
circuits.  Quantum circuits consist of quantum gates and wires. A wire
represents a qubit, and each gate, attached to one or several wires,
is a unitary operation acting on the corresponding qubits.  In this
model, a computation consists in allocating a quantum register,
applying a circuit (i.e. the list of gates, in order), followed with a
measurement to get back classical data.

The \emph{QRAM model}~\cite{knill1996conventions}
generalizes circuits: in this model,
quantum computation is
performed under the control of a classical
host. It emits a
stream of interleaved (pieces of) quantum
circuits and measurements to
the quantum co-processor. The quantum co-processor executes the
instructions while returning the results of measurements on the fly to the
classical host. In this model, the
computation is not a fixed linear list of quantum gates: the quantum
gates emitted to the quantum co-processor might depend on the results of
intermediate measurements.
Although quantum circuits and QRAM models are equivalent in terms of
expressive power, practical quantum computation is more likely to be
based on the QRAM model. For this reason, many programming languages
and their semantics are based on the QRAM model
\cite{selinger2006lambda,paykin2017qwire,green2013quipper,wecker2014liqui,omer2003structured,svore2018q,smith2016practical}.

An interesting implication of this model is that the quantum circuit
construction in the classical host can be dependent on the result of a
measurement: there is a transfer of information from the quantum
co-processor to the classical host. This feature is implemented for
example in Quipper~\cite{green2013quipper, anticoli2016towards} and
QWire~\cite{paykin2017qwire, rand2018qwire}. Following Quipper's
convention, we call this transfer \emph{dynamic lifting}: classical
information is \emph{lifted} from the quantum co-processor to the
classical host.
Some use-cases for dynamic lifting are as follows. First, 
quantum error correction typically interleaves unitaries and measurements. 
Other examples include subroutines with repeat-until-success, 
where the result of the measurement on one wire says whether the 
computation succeeded or not \cite{PhysRevLett.126.150504}, and  
measurement-based quantum computation.

The classical control over the circuit construction imposed by dynamic
lifting has not been explicitly formalized in the semantics of the
circuit construction languages using it. To illustrate this problem,
let us look at the program in Eq.~\eqref{eq:example-term}. The syntax
we use is presented in Section~\ref{s:language_type_operational_sem},
so we explain what the
program does here: The program
measures the qubit $v_c$ and obtains the updated state of the qubit
together with the resulting boolean $b$. Based on $b$, it then either
allocates a new qubit initialized by true, then free the qubit $v_c$,
or simply returns $v_c$\footnote{The program actually returns a pair
  consisting of a qubit and the unit term $*$ so that it is
  well-typed; we assume that the return type of $\text{free}$ is the
  unit-type.}. Despite this simple structure, the program does not
correspond to a circuit because of the classical control.
\begin{equation}\label{eq:example-term}
\begin{aligned}
\text{exp}\quad {::}{=}\quad
&
\textbf{let}\ \langle b, v_c \rangle = \text{meas}(v_c)\ \textbf{in}\ \quad \textbf{if}\ b\ \textbf{then}\ \langle \text{init}(\text{tt}), \text{free}(v_c) \rangle\ \textbf{else}\ \langle v_c, * \rangle
\end{aligned}
\end{equation}
In QWire, the operational semantics performs normalization for
composition and unbox operations but the classical control by dynamic
lifting is hidden in the host term within the unbox.
In Quipper, the operational semantics is encoded in Haskell's monadic
type system and captures a notion of dynamic circuit including
measurements. However, this semantics has never been fully formalized
in the context of higher-order, functional quantum languages.

Besides operational semantics, programming languages for quantum
circuits have been formalized using denotational semantics based on
density matrices \cite{paykin2017qwire} and categorical semantics
based on symmetric monoidal categories \cite{selinger2010survey,
  rios2017categorical,
  lindenhovius2018enriching,rios2017categorical,fu2020linear}, or on
the category of $C^*$-algebras
\cite{staton2015algebraic,rennela2018classical}.
However, these examples of formalization do not solve the problem in
that they either ignore the structure of circuit or keep the term with
dynamic lifting abstract.
In particular, in \cite{rios2017categorical,fu2020linear}, the authors
construct expressive categorical models for the family of circuits
---or parameterized circuits--- and linear dependent type theory,
respectively, while they do not provide semantics of dynamic lifting
explicitly.

Our goal in this paper is to find a model and formalize a semantics
for interleaved quantum circuits and dynamic lifting. The problem
rests in how to analyze the structure of the computation without
requiring the quantum co-processor to decide on the value of
measurement.
The interest of such a model and semantics is that it can serve as a
test-bed to explore properties of the language. Mixing non-duplicable
data, higher-order and circuits in a language yields a non-trivial
system, and dependent-types were for instance only added recently for
Proto-Quipper~\cite{fu2020linear}, with the use of such tools.

\subparagraph*{Contributions}
In this paper, we propose both a small step operational semantics and
a categorical semantics for a typed language ---called
Proto-Quipper-L--- extending quantum lambda calculus
\cite{selinger2006lambda} with circuit construction operators (box and
unbox) and circuit constants. The formalization extends the one of
Proto-Quipper~\cite{ross2015algebraic}: circuits are generalized to
\emph{quantum channels} enabling the formalization of the semantics of the
dynamic lifting.
A quantum computation that only consists of unitary gates
deterministically reduces to only one possible value. On the other
hand, a quantum computation with dynamic lifting might reduce to
distinct values depending on the results of the measurements.
We support this by making circuits not only lists
but trees branching over the results of measurements: we call such
objects \emph{quantum channels}. The language is then
extended with a notion of branching
terms, representing the possible choices along the computation.
We prove the usual safety properties for the language: subject
reduction, progress and termination
(Lemmas~\ref{lem:subred}, \ref{lem:prog} and~\ref{lem:term}).

Next, we propose a sound categorical model for Proto-Quipper-L.
The model is based on Proto-Quipper-M, the work of
Rios\&Selinger~\cite{rios2017categorical}. It consists of two
categories: a symmetric monoidal category abstracting the notion of
circuit, and an extension capturing classical computation and circuit
manipulation. A morphism in the former category becomes a
circuit-element in the extension. If this construction captures a
sound notion of circuit, in its abstract formulation it is not \emph{a
  priori} amenable to dynamic lifting.
To answer the issue, we propose a concrete instantiation of the model
in which dynamic lifting can be represented: we define a concrete,
symmetric monoidal closed category for representing quantum channels,
and, based on the construction proposed in~\cite{rios2017categorical},
a linear category admitting a strong monad $F$ representing the
branching side-effect associated with the measurement.
Following~\cite{moggi1991notions, valiron2008semantics}, we use the
Kleisli category $\bbMF$ to represent terms of Proto-Quipper-L.

In fact, branching monad in our categorical model corresponds to the \texttt{Circ} monad in Quipper which models non-deterministic branching 
in the level of type system. Although it is standard to use monad to 
model non-deterministic side effects, it was not clear whether
such a monadic structure could be set up on the categorical model of 
parameterized circuits by Rios\&Selinger~\cite{rios2017categorical}.
The main result of the paper is to show how to do it, using a concrete
category of quantum channels. We validate the model by showing a
soundness property (Theorem~\ref{th:sound}).

\section{Syntax, Types and Operational Semantics}
\label{s:language_type_operational_sem}

In this section, we present the syntax of a minimal lambda-calculus
for manipulating quantum channels and booleans. The language is an
extension of Proto-Quipper~\cite{ross2015algebraic}.

In Proto-Quipper, the quantum lambda calculus
presented in \cite{selinger2006lambda} is extended with circuit
operators and constants. Circuit operators give an efficient way to
construct circuits instead of having to sequentially apply all the
gates one-by-one. Specifically, two operators on circuits are added to
the language: the box operator allows us to use
quantum circuits as classical data, while the unbox operator applies a
boxed circuit to an argument (usually a structured set of qubits called
\emph{pattern}). Boxed circuits are first-class objects and can come
with useful circuit operators like reverse and control. Technically, a
circuit object in Proto-Quipper can be seen as a tuple $(p,C,M)$ where
$p$ is structured set of the input wires of a circuit $C$, matching
the input type of the circuit. $M$ is a term corresponding to the
output of the circuit. Along the reduction of $M$, the circuit $C$ is
possibly updated with new gates. The term $M$ is open, and the output
wires of $C$ are used in $M$ linearly, meaning that each output wire
appears in $M$ exactly once.

However, Proto-Quipper does not support dynamic lifting within
circuits. To extend Proto-Quipper with dynamic lifting,
we replace circuits with quantum channels and redefine the
circuit operators box and unbox over quantum channels.

\subsection{Quantum Channels}
\label{sec:qcalg}

A quantum channel is the generalization of a quantum circuit: a tree
structure where branching captures the action of measurement.
In essence, quantum channels are instances of QCAlg, defined by the
following grammar.
\[
  \text{(QCAlg)}\qquad
  Q, Q_1, Q_2\ ::=\ \epsilon(W)\
  \mid\ U(W)\ Q\ \mid\ \text{init}\ b\ w\ Q\
  \mid\ \text{meas}\ w\ Q_1\ Q_2\ \mid\ \text{free}\ w\ Q.
\]
The symbols $w$, $b$, and $W$ respectively refer to wires, booleans, and
finite sets of wires.  The channel $\epsilon(W)$ stands for the empty
computation on the qubits $W$.  $U(W)\ Q$ represent the unitary
operator $U$ acting on the qubits $W$, followed by the operations stored
in the quantum channel $Q$. In general, $U$ can range over a fixed set
of unitary operations: we write $\text{arity}(U)$ for the arity of
$U$. The operator $\text{init}\ b\ w\ Q$
creates and initializes the wire $w$ in state $b$, followed by the
channel $Q$ possibly using the newly allocated qubit.  The operator
meas represents the conditional branching on the result of a
measurement. In our interpretation, a measurement is non-destructive: the
wire being measured is still allocated and can be acted upon.
The two channels $Q_1$ and $Q_2$ stand for the two possible
branches to follow based on the measurement.  Finally, $\text{free}\ w\ Q$
frees the qubit $w$ before running $Q$.
From now on, we call the instance of QCAlg as quantum channel object.

We define a notion of validity for quantum channels: $Q$ is
\emph{valid} whenever, for instance, an init-node introduces
a non-existing wire, or whenever a free-node acts on an existing
wire. One subtlety consists in deciding what is an output
wire for a branching quantum channel. For instance, consider
\(
  Q = \text{meas}~w_1~(\text{init}~b~w_2~(\epsilon\{w_1,w_2\}))~(\epsilon\{w_1\}).
\)
This quantum channel admits as output $\{w_1,w_2\}$ on the left branch
and $\{w_1\}$ on the right branch. We formalize this notion and write
$\textbf{out}(Q)$ for a tree-structured set of outputs of $Q$: Here,
$\textbf{out}(Q)$ is $[\{w_1,w_2\},\{w_1\}]$.
We also define $\textbf{all}(Q)$ to stand for the set of
\emph{all} of the
wires appearing in $Q$, and $\textbf{in}(Q)$ for the set of input wires.
We give a formal definition of validity from the following definition 
of state of quantum channel.

\begin{definition}[State of quantum channel]\rm
  A \emph{bunch} of elements of $X$ is a binary tree where only the
  leaves are indexed, with elements of $X$. Formally, if $x$ ranges
  over $X$, a bunch is built from the grammar 
  \(
    c_1,c_2\quad{:}{:}{=}\quad x\mid [c_1,c_2].
  \)
  The ternary relation ``st'' formalizes what it means for a quantum
  channel to be valid. It is defined as the smallest relation
  satisfying the rules presented in Table~\ref{tab:valid-qc}.
  Informally, we say that a quantum channel $Q$ is \emph{valid}
  whenever there is some set of wires $V$ and a bunch of sets of wires
  $c$ such that $\textbf{st}(Q, V, c)$ is derivable. Moreover, such
  $V$ and $c$ are called input and output wires of $Q$, respectively.
\end{definition}

\begin{table}[tb]
    \small\[
    \inferrule
    { }
    {\textbf{st}(\epsilon(W),\ W,\ W)}
    \hspace{1em}
    \inferrule
    {{\begin{array}{l}W_1 \subseteq W \\ \text{arity}(U) =
       |W_1|
     \end{array}}
     \\
     \textbf{st}(Q,\ W,\ c)}
    {\textbf{st}(U(W_1)\ Q,\ W,\ c)}
    \hspace{1em}
    \inferrule
    {w \not\in W \\ \textbf{st}(Q,\ W \cup \{ w \},\ c)}
    {\textbf{st}(\text{init}\ b\ w\ Q,\ W,\ c)}
  \]
  \[
    \inferrule
    {w \in W \\ \textbf{st}(Q_1,\ W,\ c_a) \\
      \textbf{st}(Q_2,\ W,\ c_b)}
    {\textbf{st}(\text{meas}\ w\ Q_1\ Q_2,\ W,\ [c_a, c_b])}
    \hspace{2em}
    \inferrule
    {w \in W \\ \textbf{st}(Q,\ W \setminus \{ w \},\ c)}
    {\textbf{st}(\text{free}\ w\ Q,\ W,\ c)}
  \]
  \caption{Valid quantum channel}
  \label{tab:valid-qc}
\end{table}

\subsection{Syntax of the Terms}
\label{sec:terms}

Having extended the notion of circuit to the notion of quantum
channel, we turn to the question of the definition of the
language. Compared to previous Proto-Quipper
instances~\cite{ross2015algebraic,rios2017categorical,fu2020linear},
there are two main changes. The first one concerns the circuit
constant; the other one concerns the fact that one has to deal
with non-deterministic branching computations.

\begin{table}
  \small\[
\begin{aligned}
M, M_a, M_b\ ::=\ & x\ \mid\ *\ \mid\ \text{tt}\ \mid\ \text{ff}\ \mid\ (p, Q, m)\ \mid\ \lambda x.M\ \mid\ M_a M_b\ \mid\ \langle M_a, M_b \rangle\ \mid\\
& \textbf{let}\ \langle x, y \rangle = M_a\ \textbf{in}\ M_b\ \mid\
\textbf{if}\ M\ \textbf{then}\ M_a\ \textbf{else}\ M_b\ \mid\
\text{box}_P\ \mid\ \text{unbox}\\
V, V_a, V_b\,::=\,& x\,\mid\,*\,\mid\,\text{tt}\,\mid\,\text{ff}\,\mid\,\lambda x.M\,\mid\,\langle V_a, V_b \rangle\,\mid\,
(p, Q, v)\,\mid\,\text{box}_P\,\mid\,\text{unbox}\,\mid\,\text{unbox}(V)\\
p, p_a, p_b\ ::=\ & x\ \mid\ *\ \mid\ \langle p_a, p_b\rangle\\
m, m_a, m_b\ ::=\ & M \mid\ [m_a, m_b]\\
v, v_a, v_b\ ::=\ & V \mid\ [v_a, v_b]\\
A, A_a, A_b\ ::=\ & I\ \mid\ \text{bool}\ \mid\ \textbf{qubit}\ \mid\
\text{QChan}(P, A)\ \mid\ A_a \multimap A_b\ \mid\ A_a \otimes A_b\
\mid\ {!}~A
\\
P, P_a, P_b\ ::=\ & I\ \mid\ \textbf{qubit}\ \mid\ P_a \otimes P_b
\end{aligned}
\]
\caption{Proto-Quipper-L: terms, values, patterns, branching terms,
  branching values, types and pattern types.}
\label{tab:syntax}
\end{table}

We call the new language Proto-Quipper-L and define it as shown in
Table~\ref{tab:syntax}.
The constant $*$ stands for the unit term, while tt and ff stands for
the booleans true and false.  The term $(p, Q, m)$ corresponds to a
quantum channel object: $p$ is a \emph{pattern}: a
structured set of input wires of a \emph{valid}
quantum channel $Q$, and $m$ is a \emph{branching term} that will
match the branching structure of $Q$ for valid quantum channel
objects. For simplicity wire identifiers and term variables range
over the same set of names.
We then have the quantum channel operators box and unbox from
Proto-Quipper: box makes a quantum channel out of a function, while
unbox turns a quantum channel into a function.
The rest of the constructors of the language are
standard: abstraction, application, pair, let, and conditional
statements. We define a notion of value in the standard way, apart
from the fact that $\text{unbox}(V)$ is also a value (as it is a function).
Finally, branching terms and values are constructed using the
branching constructor $[-, -]$.  A term of the form $[M,N]$ represents
a computation that has probabilistically branched and that is
performing either $M$ or $N$. This is novel compared to
Proto-Quipper.
We denote the set of free variables of a term $m$ with $\text{FV}(m)$.

One could argue that the language is missing constructors for unitary
gates, qubit allocation and measurement. As in the case of
Proto-Quipper, they can be defined with the unbox and quantum channel
object. For instance, we can construct a measurement operation
inputting a qubit and outputting a boolean and the measured wire as 
$\text{meas}\ ::=\  \text{unbox}\,(x, \text{meas}~x~(\epsilon\{x\})~(\epsilon\{x\}),[\langle \text{tt}, x\rangle, \langle \text{ff}, x\rangle])$.
The tuple consists of a singleton wire name $x$, the quantum channel
(meas $x$ $\epsilon\{x\}$ $\epsilon\{x\}$), and the branching tree
$[\langle \text{tt}, x\rangle, \langle \text{ff}, x\rangle]$.  Note
that the measurement operator we wrote here returns both a qubit and a
boolean: we could discard the qubit with the use of a quantum channel
constructor ``free'' if we only wanted to output a boolean.
Similarly, we can also build the macros $\text{init}_b$ and
$\text{free}$ which respectively allocates a new qubit in state $b$
and frees a qubit, as
      $\text{init}_b\ ::=\
       \text{unbox}(*, \text{init}~b~x\,(\epsilon\{x\}), x)\,*$ and
      $\text{free}\ ::=\
      \text{unbox}\,(x, \text{free}~x~(\epsilon(\emptyset)), *)$.
We can similarly define terms for unitary application by encapsulating
the QCAlg constructors $U$ inside a quantum channel object.

\subsection{Type System}
\label{sec:types}

Types of Proto-Quipper-L are defined as in Table~\ref{tab:syntax}.
Following the standard
strategy~\cite{selinger2006lambda,ross2015algebraic,BichselBGV20} to account for the non-duplicability
brought by the quantum memory, we are using a type system based on
linear logic~\cite{girard87linear}.
Types consist in the constant types $I$, $\text{bool}$,
$\textbf{qubit}$; the function type
$A_a \multimap A_b$; the type for pairs $A_a \otimes A_b$; the
type ${!}A$ of duplicable terms of type $A$;
the type of quantum channels $\text{QChan}(P, A)$
with input of type $P$ and output of type $A$,
where $P$ refers to patterns, that is, first-order types
constructed from $\textbf{qubit}$s and tensors.

Conventionally, a typing judgment consists in a typing context, which
maps variables to types, and a term assigned with a type. However, in
Proto-Quipper-L, the term can be a branching term. Although the terms
of all branches in a branching term are assigned with the same type,
they may have different typing contexts. This is formalized in two
distinct definitions of typing judgments: regular typing judgments
$\Gamma \vdash M : A$ where where $\Gamma$ is a list of typed variables
and $M$ is a non-branching term, and \emph{branching typing judgments}
$\gamma \vdash m : A$, where $m$ is a branching term and
$\gamma$ is an branching typing context:
$\gamma ::= \Gamma\mid \gamma_1\times\gamma_2$.

\begin{table}[t]
\small
  \[
\inferrule
    { }
    {{!}\Delta,\ (x : A) \vdash x : A}
    (\text{var})
\hspace{1.5em}
\inferrule
    {{!}\Delta,\ Q \vdash M : {!}A}
    {{!}\Delta,\ Q \vdash M : A}
    (\text{d})
\hspace{1.5em}
\inferrule
    {{!}\Delta \vdash V : A \\ V\ \text{is value}}
    {{!}\Delta \vdash V : {!}A}
    (\text{p})
\hspace{1.5em}
\inferrule
    { }
    {{!}\Delta \vdash * : I}
    (\text{I})
\]
\[
\inferrule
    {{!}\Delta,\ Q,\ (x : A_a) \vdash M : A_b}
    {{!}\Delta,\ Q \vdash \lambda x.M : A_a \multimap A_b}
    (\multimap_I)
\hspace{2em}
\inferrule
    {{!}\Delta,\ Q_a \vdash M_a : A_a \multimap A_b \\ {!}\Delta,\ Q_b \vdash M_b : A_a}
    {{!}\Delta,\ Q_a,\ Q_b \vdash M_a M_b : A_b}
    (\multimap_E)
\]
\[
\inferrule
    {{!}\Delta,\ Q_a \vdash M : \text{bool} \\ {\begin{array}{l}{!}\Delta,\ Q_b \vdash M_1 : A \\ {!}\Delta,\ Q_b \vdash M_2 : A\end{array}}}
    {{!}\Delta,\ Q_a,\ Q_b \vdash \textbf{if}\ M\ \textbf{then}\ M_1\ \textbf{else}\ M_2 : A}
    (\text{if})
\hspace{1.5em}
\inferrule
   {\text{$\begin{array}{l}
          {!}\Delta,\ Q_a \vdash M_a : A_a \otimes A_b
          \\
          {!}\Delta,\ Q_b,\ (x : A_a),\ (y : A_b) \vdash M_b :A
        \end{array}$}
    }
    {{!}\Delta,\ Q_a,\ Q_b \vdash \textbf{let}\ \langle x, y \rangle = M_a\ \textbf{in}\ M_b : A}
    (\otimes_E)
\]
\[
\inferrule
    { }
    {{!}\Delta \vdash \text{tt} : \text{bool}}
    (\text{tt})
\hspace{2em}
\inferrule
    { }
    {{!}\Delta \vdash \text{ff} : \text{bool}}
    (\text{ff})
\hspace{2em}
\inferrule
    {{!}\Delta,\ Q_a \vdash M_a : A_a \\ {!}\Delta,\ Q_b \vdash M_b : A_b}
    {{!}\Delta,\ Q_a,\ Q_b \vdash \langle M_a, M_b \rangle : A_1 \otimes A_2}
    (\otimes_I)\]
\[
\inferrule
    { }
    {{!}\Delta \vdash \text{box}_P : {!}(P \multimap A) \multimap {!}\text{QChan}(P, A)}
    (\text{box})
\quad
\inferrule
    { }
    {{!}\Delta \vdash \text{unbox} : \text{QChan}(P, A) \multimap (P \multimap A)}
    (\text{unbox})
\]
\[
\inferrule
    {\gamma_a \vdash m_a : A \\ \gamma_b \vdash m_b : A}
    {\gamma_a \times \gamma_b \vdash [m_a, m_b] : A}
    (\text{b})
\hspace{3em}
\inferrule
    {p \vDash P \\ \textbf{vBind}({!}\Delta, \textbf{out}(Q), m, A)}
    {{!}\Delta \vdash (p, Q, m) : {!}\text{QChan}(P, A)}
    (\text{QChan}_I)
  \]
  \caption{Proto-Quipper-L: Typing Rules}
  \label{tab:typ-rules}
\end{table}

A judgment is valid if it can be derived from the typing rules presented in
Table~\ref{tab:typ-rules}.
The rules ensure that various constraints necessary for soundness
are satisfied. One can note that all terms constituting a
branching term share the same type\,; that valid branching typing
judgments have branching contexts and terms with the same tree
structure\,; that a quantum channel object is duplicable with type
${!}\text{QChan}$\,; that box sends a duplicable function to a
duplicable quantum channel object, and that unbox sends a quantum
channel object to a function. One can also note that only values can
be promoted to duplicable objects: this is due to the call-by-value
reduction strategy we follow.
The relation $\textbf{vBind}(!\Delta,\textbf{out}(Q),m,A)$ in the
($\text{QChan}_I$) rule ensures that one can derive typing derivations
for each term leaf of $m$ given that the output wires of the quantum
channel $Q$ is assigned with type $\textbf{qubit}$ within the
typing context.
The relation $p \vDash P$ simply states that the shapes of
$p$ and $P$ match and that the variables occurring in $p$ are pairwise
distinct.

The rules for \textbf{vBind} are found in Table~\ref{tab:vbind}. Note
  that the non-linear context ${!}\Delta$ is a list of pairs of
  variables and non-linear types. We denote by
  $\textbf{FV}({!}\Delta)$ the set of variables in ${!}\Delta$. In
  fact, the condition ($X \cap \textbf{FV}({!}\Delta) = \emptyset$) is implicitly
  assumed by the definition of the typing judgment
  (${!}\Delta,\ (x : \textbf{qubit})_{x \in X} \vdash M : A$). 

\begin{table}[tb]
  \small\[
    \begin{array}{c}
    \inferrule
  {X \cap \textbf{FV}({!}\Delta) = \emptyset \\ {!}\Delta,\ (x : \textbf{qubit})_{x \in X} \vdash M : A} {\textbf{vBind}({!}\Delta, X, M, A)}
  (\textbf{vBind}_{nb})
      \\[3ex]
      \inferrule
      {\textbf{vBind}({!}\Delta, c_a, m_a, A) \\ \textbf{vBind}({!}\Delta, c_b, m_b, A)}
      {\textbf{vBind}({!}\Delta, [c_a, c_b], [m_a, m_b], A)}
      (\textbf{vBind}_{b})
    \end{array}
    ~
    \begin{array}{c}
      \inferrule
      { }
      { * \vDash I}
      \qquad\qquad
      \inferrule
      { }
      { x \vDash \textbf{qubit} }
      \\[3ex]
      \inferrule
      { {\begin{array}{c}\forall i,~p_i \vDash P_i \\
           \textbf{FV}(p_1)\cap\textbf{FV}(p_2)=\emptyset\end{array}} }
      { \langle p_1, p_2\rangle \vDash P_1\otimes P_2 }
    \end{array}
  \]    
 \caption{Validity of binding in quantum channel constant}
 \label{tab:vbind}
\end{table}

\begin{example}
  In Section~\ref{sec:terms} we defined three macros: meas, free and
  init${}_b$. We can type meas
  with $!(\textbf{qubit}\multimap (\text{bool}\otimes\textbf{qubit}))$ and
  free with $!(\textbf{qubit}\multimap I)$. For $\text{init}_b$, note
  that because there is a final argument ``$*$'', it is really an
  application and we can therefore only type it with $\textbf{qubit}$
  and not $!\textbf{qubit}$: this is expected, as we don't want to
  be able to construct duplicable qubits.
  With these types, we can now type the term exp in
  Eq~\eqref{eq:example-term} of Section~\ref{s:introduction}: we can
  derive the judgment
  $v_c:\textbf{qubit}\vdash\text{exp}:\textbf{qubit}\otimes I$.
\end{example}

\begin{remark}\label{rem:basic-type}
  In general, there can be more than one typing derivation for a
  typing judgment but, for the types $I$, \textbf{qubit} or bool, there is a unique
  typing derivation when the term is a value. We call these types
  \emph{basic types}.
\end{remark}

\subsection{Operational Semantics}
\label{s:operational_semantics}

The computational model we have in mind for the language is a
reduction-based semantics
specialized to circuit construction: the operational
semantics is modeling an I/O side-effect, where gates are
emitted and buffered in a quantum channel.
Based on Proto-Quipper~\cite{ross2015algebraic}, the operational semantics we
describe therefore updates a \emph{configuration} consisting of a pair
$(Q,m)$: a buffered QCAlg object and a branching term. The term $m$ is
reduced up to a value representing the final state of the
computation. Along the computation, quantum gates might be emitted to
the co-processor: the quantum channel $Q$ keeps track of these.
One can notice that a configuration corresponds to a quantum channel
constant without the input wires, where there is a minor relaxation on
the linearity of the output wires of $Q$ in $m$ which will be
recovered when we define well-typed configuration.

\begin{definition}\rm
  A \emph{circuit-buffering configuration} is a pair $(Q,m)$ as
  described above. It is said to
  be \emph{valid} whenever $Q$ is valid, $Q$ and $m$ share the same
  tree-structure, and whenever output wires of $Q$ corresponds to free
  variables of $m$ (following the tree-structure). So for instance,
  $V \subseteq \textbf{FV}(M)$ implies the validity of
  $(\epsilon(V),\ M)$, and whenever $(Q_1,\ m_a)$ and $(Q_2,\ m_b)$
  are valid so is $(\text{meas}\ w\ Q_1\ Q_2,\ [m_a, m_b])$.
\end{definition}

\begin{remark}
  In order to define the reduction rules, we need to be able to extend
  a configuration with new wires. For instance, let us consider the
  term
  \( (\epsilon\{x,y\}, \textbf{if}\ (N\,x)\ \textbf{then}\ y\
  \textbf{else}\ y) \)
  with $N$ some term not containing $y$.  Evaluating this
  configuration requires to first evaluate $N\,x$ and possibly append
  a few gates to $\epsilon\{x,y\}$. However, this can be factorized
  as first evaluating $(\epsilon\{x\}, N\,x)$ to $(Q, V)$ and then
  adding back the wire $y$ to the resulting quantum channel $Q$. We
  therefore define an operator $\textbf{extend}$ taking a quantum
  channel and a set of wire names, adding them as unused wires to
  the quantum channel.
\end{remark}

The reduction rules for Proto-Quipper-L are defined in
Table~\ref{tab:red-rules}. (See Section~\ref{app:definition-red} for more details.)
Rules (a.x) always hold ($b$ ranges over $\{\text{tt},\text{ff}\}$).  
In Rules (b.1), $p$ is a pattern of same shape as $P$ made from
  dynamically allocated fresh variables. In Rule (b.2), $p$ and $V$
  have the same shape, and $\sigma$ is a substitution mapping $p$ to
  $V$.
  Provided that $(Q, m) \xrightarrow{} (Q', m')$, we have
  $(\epsilon(\emptyset), (p, Q, m)) \xrightarrow{}
  (\epsilon(\emptyset), (p, Q', m'))$.
  Provided that we have that
  $(\epsilon(W_M),\ M) \xrightarrow{} (Q,\ m)$, that
  $\textbf{all}(Q) \cap W_N = \emptyset$ and that
  $\textbf{all}(Q) \cap W_V = \emptyset$, the class of rules (c) apply.
  There, $C[-]$ ranges over $[-]N$, $V[-]$, $\langle [-],N\rangle$,
  $\langle V,[0]\rangle$,
  $\text{if}\,[-]\text{then}\,M_a\,\text{else}\,M_b$ and
  $\text{let}\,\langle x,y\rangle=[-]\,\text{in}\,N$.
  We use syntactic sugar for
  combining terms and branching terms, as in $C[m]$.
  It corresponds to the term constructor applied to each leaf of 
  $m$, for instance: for $m = [[N_1,N_2],N_3]$, 
  $C[m] := [[C[N_1],C[N_2]],C[N_3]]$.
  In Rules (d.x), $Q$ stands for $\text{meas}\ w\ Q_1\ Q_2$ and $Q'$
  for $\text{meas}\ w\ Q_3\ Q_4$.  These rules apply whenever 
  $(Q_1,\ m_a) \xrightarrow{} (Q_3,\ m_c)$ and
  $(Q_2,\ m_b) \xrightarrow{} (Q_4,\ m_d)$.
  In (d.3), $G$ ranges over $U(W)$, $\text{init}~b~w$ and
  $\text{free}~w$.

\begin{table}[t]
  {\scalebox{.73}{\begin{minipage}{4.1in}
  \[
    \begin{aligned}
      \text{(a.1)}&&(\epsilon(W), (\lambda x.M) V) &\xrightarrow{} (\epsilon(W), M[V / x])
      \\
      \text{(a.2)}&&(\epsilon(W), \textbf{let}\ \langle x, y \rangle = \langle V, U \rangle\ \textbf{in}\ M) &\xrightarrow{} (\epsilon(W), M[V / x, U / y])
      \\
      \text{(a.3)}&&(\epsilon(W), \textbf{if}\ \text{b}\ \textbf{then}\ M_{\text{tt}}\
      \textbf{else}\ M_{\text{ff}}) &\xrightarrow{} (\epsilon(W),
      M_{b})
      \\
      \text{(b.1)}&&
      (\epsilon(\emptyset), \text{box}_P\,V) &\xrightarrow{}
      (\epsilon(\emptyset), (p, \epsilon(\text{FV}(p)), V p))
      \\
      \text{(b.2)}&&
      (\epsilon(\textbf{FV}(V)), (\text{unbox}(p,\ Q,\ u))V) &\xrightarrow{} (\sigma(Q), \sigma(u))\\~
    \end{aligned}
  \]
\end{minipage}}}
\hspace{-6ex}
{\scalebox{.73}{\begin{minipage}{3.5in}
\[
  \begin{aligned}
    (\epsilon(W_{C[M]}),\ C[M])&\xrightarrow{}
    (\textbf{extend}(Q,W_{C[-]}),\ C[m]) & \text{(c)}
    \\
    (Q,\ [m_a, m_b]) &\xrightarrow{}
    (Q',\ [m_c, m_d]) & \text{(d.1)}
    \\
    (Q,\ [m_a, v]) &\xrightarrow{}
    (Q',\ [m_c, v]) & \text{(d.2)}
    \\
    (Q,\ [v, m_b]) &\xrightarrow{}
    (Q',\ [v, m_d]) & \text{(d.3)}
    \\
    (G\ Q_1,\ m_a) &\xrightarrow{} (G\ Q_3,\ m_c)  & \text{(d.3)}\\~
  \end{aligned}
\]
\end{minipage}}}
  \caption{Reduction rules for operational semantics.}
  \label{tab:red-rules}
\end{table}

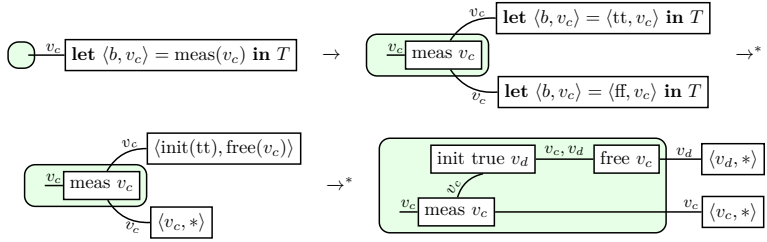
\begin{figure}[t]
  \scalebox{.7}{
    \begin{minipage}{6in}
  \tikzstyle{operator} = [draw,fill=white,minimum size=1.5em] 
  \tikzstyle{measure} = [draw,fill=white,minimum size=1.5em,path picture = 
    {\draw[black] ([shift={(.1,.25)}]path picture bounding box.south west) to[bend left=50] ([shift={(-.1,.25)}]path picture bounding box.south east);
    \draw[black,-latex] ([shift={(0,.1)}]path picture bounding box.south) -- ([shift={(.2,-.1)}]path picture bounding box.north);}]
  \tikzstyle{phase} = [fill,shape=circle,minimum size=5pt,inner sep=0pt]
  \tikzstyle{surround} = [fill=green!10,thick,draw=black,rounded corners=2mm]
  \tikzstyle{box} = [draw,minimum width=2.3em,fill=white,minimum height=7em]
  \centerline{
    \begin{tikzpicture}[baseline=-0.5ex,thick]
    %
    \draw (0,0) node[] (qs) {};
    \draw (3,0) node[rectangle, draw] (qt) {$\textbf{let}\ \langle b, v_c \rangle = \text{meas}(v_c)\ \textbf{in}\ T$};
    %
    \path[every node/.style={font=\sffamily\small, inner sep=1pt}]
    (qs) edge [-, sloped, pos=0.7, above] node {$v_c$} (qt);
    %
    \begin{pgfonlayer}{background} 
    \node[surround] (background) [fit = (qs)] {};
    \end{pgfonlayer}
    \end{tikzpicture}
  \quad $\xrightarrow{}$ \quad
    \begin{tikzpicture}[baseline=-0.5ex,thick]
    %
    \draw (0,0) node[] (qs) {};
    \draw (1.2,0) node[operator] (qm) {$\text{meas}\ v_c$};
    \draw (4.2,0.7) node[rectangle, draw] (qt1) {$\textbf{let}\ \langle b, v_c \rangle = \langle \text{tt}, v_c \rangle\ \textbf{in}\ T$};
    \draw (4.2,-0.7) node[rectangle, draw] (qt2) {$\textbf{let}\ \langle b, v_c \rangle = \langle \text{ff}, v_c \rangle\ \textbf{in}\ T$};
    %
    \path[every node/.style={font=\sffamily\small, inner sep=1pt}]
    (qs) edge [-, sloped, above] node {$v_c$} (qm)
    (qm) [bend left=30] edge [-, sloped, pos=0.7, above] node {$v_c$} (qt1.west)
    (qm) [bend right=30] edge [-, sloped, pos=0.7, below] node {$v_c$} (qt2.west);
    %
    \begin{pgfonlayer}{background} 
    \node[surround] (background) [fit = (qs) (qm)] {};
    \end{pgfonlayer}
    \end{tikzpicture}
  \quad $\xrightarrow{}^*$ \quad
  }
  \bigskip
  \centerline{
    \begin{tikzpicture}[baseline=-0.5ex,thick]
    %
    \draw (0,0) node[] (qs) {};
    \draw (1.2,0) node[operator] (qm) {$\text{meas}\ v_c$};
    \draw (3.5,0.7) node[rectangle, draw] (qt1) {$\langle \text{init}(\text{tt}), \text{free}(v_c) \rangle$};
    \draw (2.7,-0.7) node[rectangle, draw] (qt2) {$\langle v_c, * \rangle$};
    %
    \path[every node/.style={font=\sffamily\small, inner sep=1pt}]
    (qs) edge [-, sloped, above] node {$v_c$} (qm)
    (qm) [bend left=30] edge [-, sloped, pos=0.7, above] node {$v_c$} (qt1.west)
    (qm) [bend right=30] edge [-, sloped, pos=0.7, below] node {$v_c$} (qt2.west);
    %
    \begin{pgfonlayer}{background} 
    \node[surround] (background) [fit = (qs) (qm)] {};
    \end{pgfonlayer}
    \end{tikzpicture}
  \quad $\xrightarrow{}^*$ \quad
    \begin{tikzpicture}[baseline=-0.5ex,thick]
    %
    \draw (0,-0.5) node[] (qs) {};
    \draw (1.2,-0.5) node[operator] (qm) {$\text{meas}\ v_c$};
    \draw (1.7,0.5) node[operator] (qi) {$\text{init}\ \text{true}\ v_d$};
    \draw (4.4,0.5) node[operator] (qf) {$\text{free}\ v_c$};
    \draw (6.4,0.5) node[rectangle, draw] (qt1) {$\langle v_d, * \rangle$};
    \draw (6.4,-0.5) node[rectangle, draw] (qt2) {$\langle v_c, * \rangle$};
    %
    \path[every node/.style={font=\sffamily\small, inner sep=1pt}]
    (qs) edge [-, sloped, above] node {$v_c$} (qm)
    (qm)  edge [-, sloped, bend left=30, pos=0.2, above] node {$v_c$} (qi.south)
    (qi) edge [-, sloped, above] node {$v_c, v_d$} (qf)
    (qf) edge [-, sloped, pos=0.6, above] node {$v_d$} (qt1)
    (qm) edge [-, sloped, pos=0.95, above] node {$v_c$} (qt2);
    %
    \begin{pgfonlayer}{background} 
    \node[surround] (background) [fit = (qs) (qm) (qi) (qf)] {};
    \end{pgfonlayer}
    \end{tikzpicture}
  }
\end{minipage}}
\caption{Reduction of the term of Eq~\eqref{eq:example-term}}
  \label{fig:abs_reduction_example}
\end{figure}

\begin{example}
\label{sec:example-red} 
As an example, we show the reduction of the term shown in
Eq. \eqref{eq:example-term}.
For convenience, we define $T$ as
$\textbf{if}\ b\ \textbf{then}\ \langle \text{init}(\text{tt}),
\text{free}(v_c) \rangle\ \textbf{else}\ \langle v_c, *
\rangle$. Figure \ref{fig:abs_reduction_example} shows the reduction
of the term. (check Section~\ref{app:example-red} for more details).
We use a graphical representation for configuration. A green box
represents a quantum channel whose leaves are linked to square-boxed terms.
The edges represent bundles of wires, which can contain
multiple wires and can be empty.

In the first line, the measurement in the term is reduced
by the structural rule for $let$ and the reduction rule for
measurement creating a branching term. Then, each term at a leaf of
the tree is reduced into the left-most configuration of the second
line. Note how classical computation can happen inside the leaves. The
second line of the figure shows the application of initialization and
free operation. In particular, note how the tree expands as the
computation progresses.
\end{example}

\subsection{Type safety for Proto-Quipper-L}

In order to state the type safety theorem, we need to extend typing
derivations to configurations. We write ${!}\Delta \vdash (Q, m) : A$
whenever
$\textbf{vBind}(!\Delta,\textbf{out}(Q),m,A)$ and $Q$ is valid.
Note that the definition implies that the output wires of the quantum
channel correspond to the linear variables of type $\textbf{qubit}$
in the context of the typing derivation.
In any case, we can now state type safety for
Proto-Quipper-L, as follows.

\begin{lemma}[Subject reduction]\label{lem:subred}
  For any configurations $(Q_1, m_1)$ and $(Q_2, m_2)$ such that
  $(Q_1, m_1) \xrightarrow{} (Q_2, m_2)$, if $\vdash (Q_1, m_1) : A$,
  then $\vdash (Q_2, m_2) : A$.\qed
\end{lemma}

\begin{lemma}[Progress]\label{lem:prog}
  If $(\vdash (Q_1, m_1) : A)$, then either there exists $(Q_2, m_2)$
  such that $(Q_1, m_1) \xrightarrow{} (Q_2, m_2)$ or $m_1$ is a
  branching value.\qed
\end{lemma}

\begin{lemma}[Termination]\label{lem:term}
  Given a well-typed configuration $\vdash (Q,m):A$, any reduction
  sequence starting with $(Q,m)$ is terminating.\qed
\end{lemma}

\section{Categorical semantics}\label{s:categorical_semantics}

In this section, we turn to the question of developing a categorical
semantics for Proto-Quipper-L.
The categorical semantics of circuit-description
languages and Proto-Quipper in particular originates from
Rios\&Selinger~\cite{rios2017categorical}. They developed a model
parametrized by a symmetric monoidal category $M$. In their model
one can therefore interpret higher-order
circuit-description languages, and several
extensions of the
semantics~\cite{fu2020linear,lindenhovius2018enriching} have
been discussed. However, none of them were shown to be
able to capture dynamic lifting: the
possibility to change behavior depending on the result of a
measurement. 

\subparagraph*{Our proposal}
What we propose in this paper is a concrete, symmetric monoidal
category $M$ such that applying Rios\&Selinger's construction gives us
also access to an interpretation of dynamic lifting.
The model we propose follows Moggi's categorical
interpretation of side effect~\cite{moggi1991notions} and models the
action of measurement using a (strong) monad. Our semantics is
therefore based on:
(1) A category of \emph{diagrams}, serving as graph-like abstractions
  of quantum channels. This category is compact-closed and features
  products: it matches the requirements of the base category $\bM$ in
  Rios\&Selinger's work. This category is
  discussed in Section~\ref{sec:diag}.
(2) The category $\bbM$, extending
  $\bM$ with the same procedure as Rios\&Selinger. This
  category is the category of \emph{values}, following Moggi's
  computational interpretation. It is presented in
  Section~\ref{sec:coprod-compl}.
(3) A strong monad on $\bbM$ that we denote with
  $F$.
  This monad encapsulates \emph{computations}
  involving measurements: a general term of Proto-Quipper-L is
  therefore interpreted inside the Kleisli category
  $\bbMF$: This is the main novelty compared to other models of
  Proto-Quipper-like
  languages~\cite{rios2017categorical,fu2020linear,lindenhovius2018enriching},
  and the critical reason for the possibility to interpret dynamic
  lifting.  This is discussed in Section~\ref{sec:kleisli}.

Finally, we discuss the soundness of the model and
presents a few examples. For sake of space, the presentation of the
definitions and results is only kept to a minimum: more information is
available in the appendix.

\subsection{Categories of Diagrams}
\label{sec:diag}

In this section, we aim at building a category of quantum channels.
We first define a graph-based language: we call the corresponding
terms \emph{diagrams} to distinguish them from the terms of QCAlg of
Section~\ref{sec:qcalg}: these are directed graphs with edges labeled
with
\emph{marks}. We then build the category $\bM$ out of these
terms.

\subparagraph*{Marks}
Formally, we define \emph{marks} with the
grammar
\(
M\ ::=\ q\ \mid\ M \otimes M
\ \mid\ \boxplus_{i \in X} M_i\ \mid\ M^{\bot},
\)
where $X$ ranges over the class of sets, and is subject to the
equivalence relation defined as
$\boxplus_{i \in I} \boxplus_{j \in J} M_{(i,j)} = \boxplus_{j \in J}
\boxplus_{i \in I} M_{(i,j)} $;
 $(M_1 \otimes M_2)^{\bot} = M_1^{\bot} \otimes M_2^{\bot}$;
 $(\boxplus_{i \in I} M_i)^{\bot} = \boxplus_{i \in I} (M_i^{\bot})$;
 $(M^{\bot})^{\bot} = M$;
 $\boxplus_{l \in L} \boxplus_{x \in l} M_{(l, x)} = \boxplus_{x \in
   l_1 ++ \cdots ++ l_n} M_{(l, x)}$, whenever $L = [l_1, \ldots,
 l_n]$.
 Note that $\boxplus_{i\in\emptyset}M_i$ acts as a unit for
 $\boxplus$: we denote it with $I$.  If $A=[A_1\ldots A_n]$ is a list
 of marks, we use the notation $A^{\otimes}$ for
 $A_1\otimes\cdots\otimes A_n$. We also use a binary notation for
 $\boxplus$ when the indexed set contains 2 elements:
 $\boxplus_{x\in\{a,b\}}A_x = A_a\boxplus A_b$.

\begin{remark}
  Box node is a way of representing additive connectives of
  intuitionistic linear logic. It can be considered as a set of
  different proofs depending on the choice made for the additive
  connective. Note that we are following the convention of linear
  logic for $(-)^\perp$, where the $(-)^\perp$ operator is not changing the
  order of tensors.
\end{remark}

\subparagraph*{Diagrams}
A diagram is a (possibly infinite)
directed graph with edges indexed with marks and built
from \emph{elementary nodes} and \emph{boxes}. A
diagram is not necessarily a connected graph. By graphical convention,
all edges are flowing upward: a diagram is therefore acyclic.

\begin{figure}[t]
  \begin{subfigure}{2.3in}
    \includegraphics[width=2.2in]{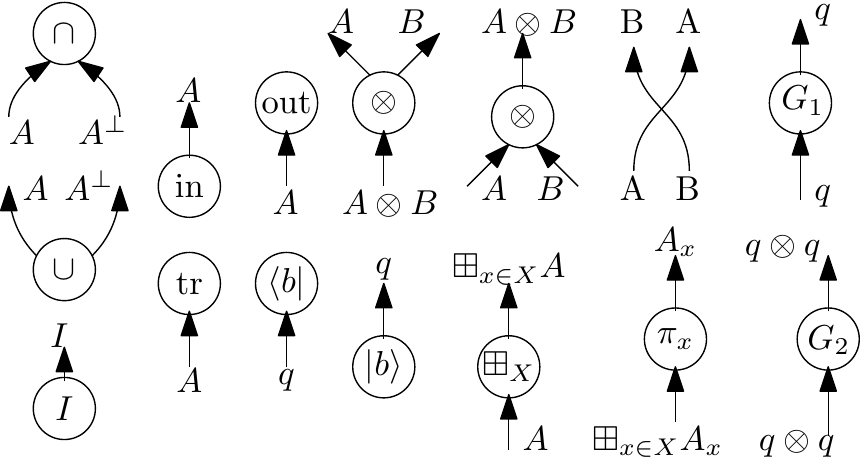}
    \caption{Elementary nodes}
    \label{fig:elem-nodes}
  \end{subfigure}
  \hfill
  \begin{subfigure}{1.8in}
     \includegraphics[width=1.7in]{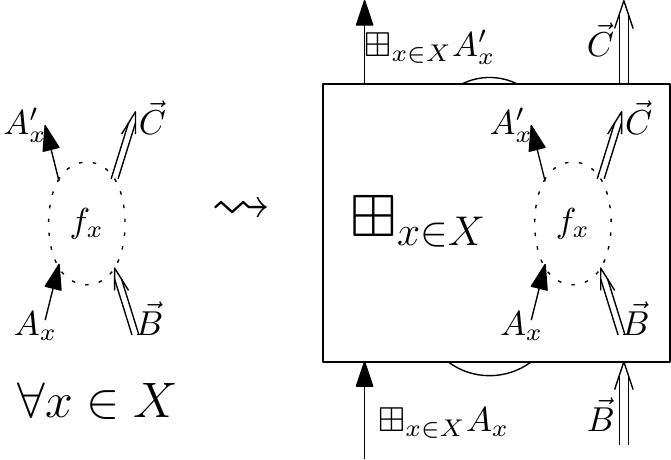}
    \caption{Box node}
    \label{fig:box-node}
  \end{subfigure}
  \hfill
  \begin{subfigure}{0.65in}
  \includegraphics[width=0.6in]{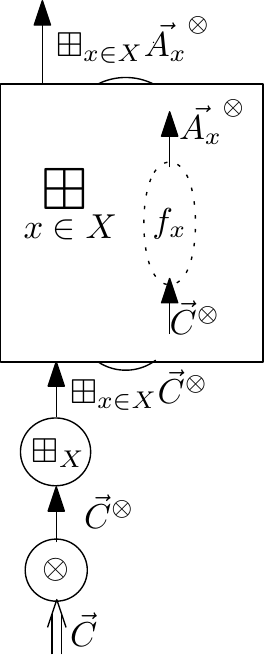}
    \caption{Product}
    \label{fig:prod-diag}
  \end{subfigure}
  \caption{Diagram Nodes and Product}
  \label{fig:nodes}
\end{figure}

Elementary nodes make the basic building blocks of diagrams:
they are shown on Figure~\ref{fig:elem-nodes}. As we
work with directed graphs, each edge connected to a node is either an
input or an output for that node. There are several kinds of elementary
nodes: the structural nodes for capturing the compact closed
structure: \rbox{\cup}, \rbox{\cap}, \rbox{\otimes}, \rbox{I} and the
swap-node (also written \rbox{\sigma}); the structural nodes for
handling the product: \rbox{\boxplus} for the diagonal map and
\rbox{\pi} for the projection; the
structural nodes for pointing inputs \rbox{\text{in}}  and outputs
\rbox{\text{out}} of diagrams; the nodes specifically for quantum:
\rbox{|b\rangle} and \rbox{\langle b|}, with $b$ ranging over
booleans, where the former stands for initialization and the latter
for projection onto the corresponding basis, \rbox{\text{tr}} for
representing tracing (also useful for products),
\rbox{G_1} for unary unitary gates and
\rbox{G_2} for binary gates.
Note that the nodes allows more expressivity than what we need: for
instance, \rbox{\text{tr}} and \rbox{\langle b|} are
indistinguishable. We nonetheless keep them in order to draw attention
on the correspondence with quantum computation and an obvious
mapping to completely positive maps.
For the sake of legibility,
we do not draw in and out nodes unless necessary.

Presented in Figure~\ref{fig:box-node},
a box-node is built from a family of diagrams. They should all
share the same
input and output marks except for one pair of input/output
(represented on the left of the box-node). As
a node, box-node has the same input and output marks as
its contained diagrams except that the left-most marks: these
are the $\boxplus$ of all left-most marks of the family.
We represent juxtaposition of edges as a double arrows.
This node is the last piece needed for representing products.

\subparagraph*{Equivalence relation on diagrams}
We define an equivalence relation on diagrams. The equivalence is
given with local rules that can be extended to larger diagrams
coherently: subgraphs can be rewritten inside a larger graph. These
rules exactly capture what is needed for the categorical semantics to
work. For instance, we include all of the rules for compact closed
categories~\cite{selinger2010survey}. We also for instance need the
fact that the $\rbox{\pi}$-node acts as a projection over box-nodes. The
complete list can be found in the appendix.

\subparagraph*{Category of Diagrams}
Based on the definition of diagrams, we define the category of diagrams
$\bM$: object are lists of marks $[A_1, \ldots, A_n]$, and a morphism
$[A_1, \ldots, A_n] \to [B_1, \ldots, B_m]$ is a diagram
with \rbox{\text{in}}-nodes
of marks $A_i$ and \rbox{\text{out}}-nodes of marks $B_i$, modulo the
equivalence relation defined on diagrams. We use the notation
$\vec{A}$ for the list of the $A_i$'s.
An identity morphism is a diagram consisting of a bunch of simple edges
connecting \rbox{\text{in}} and \rbox{\text{out}} nodes.
Composition consists in identifying 
\rbox{\text{out}} and \rbox{\text{in}} nodes of diagrams.
The category $\bM$ is symmetric monoidal: The unit is $I =
[]$, the empty list, and the monoidal
structure is given with
$\otimes : \bM \times \bM \to \bM$ defined as
$[A_1, \ldots, A_n] \otimes [B_1, \ldots, B_m] = [A_1, \ldots, A_n,
  B_1, \ldots, B_m]$, and $f \otimes g$ the juxtaposition of diagrams.
As for standard graphical representation of symmetric monoidal
structure~\cite{selinger2010survey}, the associativity, unit laws and
symmetry of the tensor product follow their graphical conventions.
Finally, the operation on marks $(-)^\bot$ lifts to a contravariant
functor, giving a compact-closed structure to $\bM$. It then
admits an internal hom: $\vec{A} \multimap \vec{B}$ can be defined as
$[A_1, \ldots, A_n] \multimap [B_1, \ldots, B_m] = [A_1^{\bot},
\ldots, A_n^{\bot}, B_1, \ldots, B_m]$. Thanks to the \rbox{\pi}-nodes
and
the corresponding diagram equivalence rules, the category
$\bM$ also has products:
 for any family of objects $\{ \vec{A_x} \mid x \in X \}$ indexed by a
 set $X$, let
 $\times_{x \in X} \vec{A_x} = [\boxplus_{x \in X}
 \vec{A_x}^{\otimes}]$ be the product of the family of objects.
Then, the family of projections
  $\pi_x : \times_{x \in X} A_x \to A_x$ is given by the $\pi$-node.
Finally, for any family of maps $\{ f_x : C\to A_x\}_{x \in X}$, the morphisms
 $\langle f_x \rangle : C \to \times_{x \in X} A_x$ is the 
 diagram presented in Figure~\ref{fig:prod-diag}. As an abuse of notation we
 use one \rbox{\otimes} for tensoring several wires at once.

 \begin{remark}
  The category of diagrams is strongly inspired from proof nets:
  tensor nodes correspond to multiplicative connectives while boxes
  correspond to additive connectives.
\end{remark}

\begin{figure}[bt]
  \begin{subfigure}{2.5in}
    \includegraphics[width=2.4in]{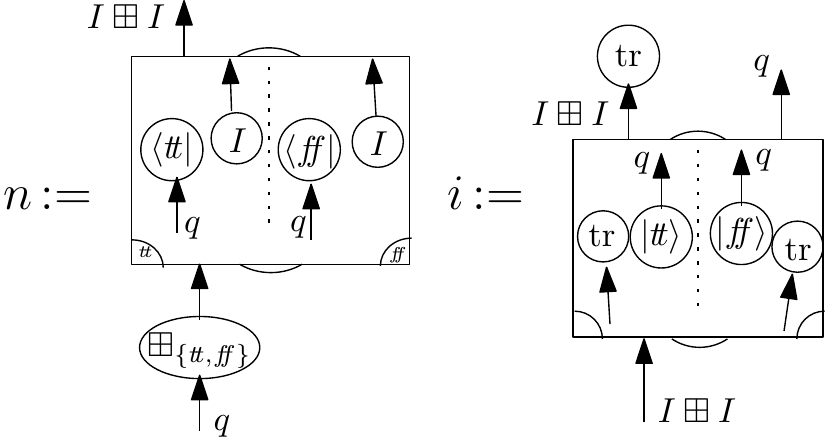}
    \caption{Morphisms in $\bM$}
    \label{fig:morph-bM}
  \end{subfigure}
  \hfill
  \begin{subfigure}{2.5in}
    \includegraphics[width=2.4in]{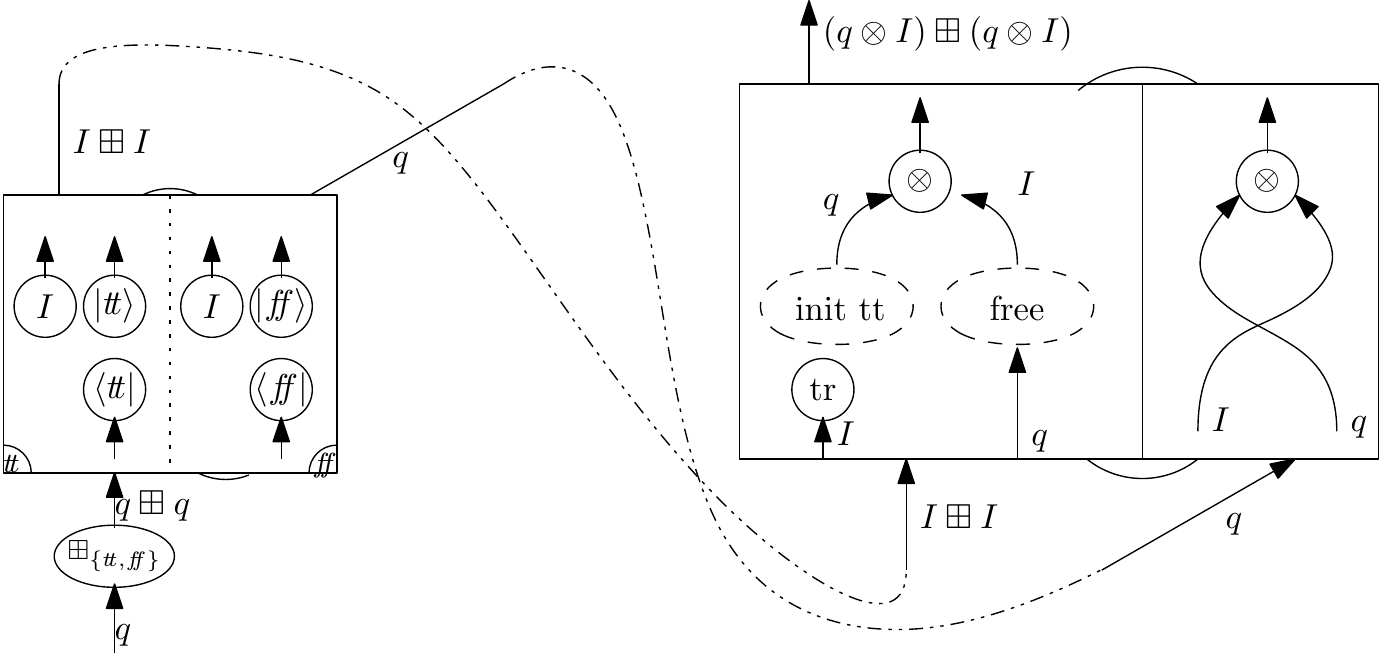}
    \caption{Figure for Example~\ref{ex:blah}}
    \label{fig:ex-den}
  \end{subfigure}
  \caption{Examples of Morphisms}
\end{figure}
 
\subparagraph*{Examples of morphisms in $\bM$}
Lastly, we show in Figure~\ref{fig:morph-bM} two interesting morphisms in
the category $\bM$. The morphism $n:[q]\to[I\boxplus I]$ corresponds
to the measure:
in each branch we perform a projection, and we keep in
the output the information of where we were. Note that the semantics
does not state what is doing $\langle t\!t|$: what is important is to
(1) ``remove'' the $q$-wire, and (2) keep as information if we are on
the ``true'' or the ``false'' part. The morphism $i$ corresponds to
qubit creation: it takes a boolean $I\boxplus I$, initializes a qubit
depending on its state and ``forgets''
the boolean. As a last example we can build
the injections $I\to I\boxplus I$ in a similar way to
$n$: first a \rbox{\boxplus}-node, followed with a box-node
where we trace
out the component we do not need.

\begin{remark}
  The object $[I\boxplus I]$ corresponds to the bit-type in Quipper or
  in Proto-Quipper, corresponding to boolean values within the quantum
  co-processor, and manipulated with circuits in Quipper. For
  simplicity we did not include such a bit-type in the language, but
  it does exist in the model.
\end{remark}

\begin{definition}[Intepreting QCAlg terms]\label{def:denot-qcalg}\rm
  Let us use the notation $q^{\otimes n}$ to represent a list
  $[q,...q]$ of size $n$.  A QCAlg-term $Q$ can be interpreted as a
  $\bM$-morphism $\denot{Q}:A\to B$, where
  $A=q^{\otimes\textbf{in}(Q)}$ and $B$ of tree-shape for instance
  $(q^{\otimes n_1}\boxplus q^{\otimes n_2})\boxplus q^{\otimes n_3}$,
  following the tree-shape of $\text{out}(Q)$. The $\bM$-morphism
  $\denot{Q}$ is then defined by induction, using the idea presented
  above: initialization and unitary gates are simply composed, and the
  branches of a meas operation are encapsulated inside box-nodes.
\end{definition}

\subsection{Coproduct completion}
\label{sec:coprod-compl}

Coproduct completion allows us to define families of circuits
\cite{rios2017categorical, fu2020linear}: the categorical structure
clearly separate what is \emph{purely quantum} and what is
\emph{parameter} to the computation: we have \emph{parametric}
families of quantum channels.
Formally, this is done using the coproduct
completion $\bbM$ of $\bM$.
In this completion, an object corresponds to a pair
$(X, (A_x)_{x \in X})$ where $X$ is a set and $A_x$ is an object of
$\bM$ for each $x \in X$: This should be understood as a
\emph{parametric} families of objects of $\bM$.
A morphism from $(X, (A_x))$ to $(Y, (B_y))$ corresponds to a pair
$(f_0, (f_x)_{x \in X})$ where $f_0 : X \to Y$ is a set function and
$f_x : A_x \to B_{f_0(x)}$ is a morphism in $\bM$ for each
$x \in X$. Intuitively, to each choice of parameter $x$ we have a
$\bM$-morphisms $A_x\to A_{f_0(x)}$.
Composition is defined with
$(g_0, (g_y)) \circ (f_0, (f_x)) = (g_0 \circ f_0, (g_{f_0(x)} \circ
f_x))$ where $(g_0, (g_y)) : (Y, (B_y)) \to (Z, (C_x))$ and
$(f_0, (f_x)) : (X, (A_x)) \to (Y, (B_y))$ are morphisms in
$\bbM$, while the identity is
$\textbf{id}_A = (\textbf{id}_X, (\textbf{id}_{A_x}))$ for an object
$A = (X, (A_x))$.

According to Rios\&Selinger~\cite{rios2017categorical}, the category
$\bbM$ is symmetric monoidal closed, and features
products and co-products. In particular, the monoidal unit is
$(\{ \emptyset \}, (I))$ (where $\emptyset$ stands for the only
representative of the singleton-set), and when
$A = (X, (A_x))$ and $B = (Y, (B_y))$, the tensor on objects is
$A \otimes B = (X \times Y, (A_x \otimes B_y)_{(x,y)})$  and
the internal hom is
$A \multimap B = (X \to Y, (C_f)_{f \in X \to Y})$ ($X \to Y$ is
the set of all set-functions from $X$ to $Y$ and $C_f$ refers to the
product $\boxplus_{x \in X} (A_x \multimap B_{f(x)})$ of internal homs
in $\bM$). Note that the product is defined by $\boxplus$ in
the case of the category of diagrams.
Also note that compared to~\cite{rios2017categorical}, we can
capitalize on the concrete structure of the category for the proofs
involving the coproduct completion. For instance, the associativity is
trivial in our category $\bM$.

Finally, in order to model the type operator ``$!$'', Rios\&Selinger
rely on Benton's linear/non-linear model \cite{benton1994mixed}, based
on an adjunction between a symmetric monoidal closed category and a
cartesian closed category. In our case, as
in~\cite{rios2017categorical} the adjunction is built
between the SMCC $\bbM$ and the cartesian closed
category $\mathbf{Set}$. The two functors of the adjunction are
$p : \mathbf{Set} \to \bbM$, defined on objects as
$p(X) = (X, (I)_X)$, and
$b : \bbM \to \mathbf{Set}$, defined on objects as
$b(X, (A_x)) = \sum_{x \in X} \bM(I, A_x)$ where
$\bM(I, A_x)$ is the set of morphisms between the objects $I$
and $A_x$ of the category $\bM$ and
$\sum_{x \in X} \bM(I, A_x)$ is the disjoint union of all
such sets over $X$ . From the adjunction, one can then construct a
comonad ``$!$'' defined as $! = p \circ b$.

\begin{remark}\label{rem:bool}
  In $\bbM$ there are two classes of interesting objects. The
  \emph{parameters} are objects of the form $(X,(I)_{x\in X})$: the
  family consists in trivial objects of $\bM$, and the only
  information is given by\ldots the parameter. The \emph{state} object
  is the dual: the parameter is trivial and the family is of size 1
  with only one object of $\bM$. It is then of the form
  $(\{\emptyset\},(A))$. One therefore has two booleans: a
  parameter boolean
  $b_p = (\{t\!t,f\!\!f\},(I)_{\{t\!t,f\!\!f\}})$ and the
  state boolean $b_s = (\{\emptyset\},(I\boxplus I))$
  living in $\bM$.
\end{remark}

\subsection{Monad for Branching Computation}

According to Rios\&Selinger, the category $\bbM$
together with the structure sketched in Section~\ref{sec:coprod-compl}
forms a model of Proto-Quipper-M.
We shall now see how our concrete construction can also
support dynamic lifting, therefore forming a model of Proto-Quipper-L.

The main problem consists in \emph{lifting} a branching sitting inside
a quantum channel ---i.e. inside the category $\bM$--- to turn it into
a coproduct on which one can act upon in the classical world,
represented by the category $\bbM$: as in Remark~\ref{rem:bool}, we
need to lift a state-boolean into a parameter-boolean.
Our strategy consists in defining a strong monad $(F, \mu, \eta, t)$
to capture the action of retrieving such a branching: a term featuring
measurement (and dynamic lifting) is therefore represented within the
Kleisli category $\bbMF$, following
Moggi's~\cite{moggi1991notions} view on side-effects.

The functor $F : \bbM \to \bbM$
is defined as follows. For an object $A = (X, (A_x))$, we define
$F(A) = (\text{mset}(X), ([\boxplus_{x \in l} A_x^{\otimes}])_{l \in
  \text{mset}(X)})$, where $\text{mset}(X)$ is the set of multisets
of $X$, while for a morphism $f = (f_0, (f_x)) : A \to B$ we set
\(
F(f) = (g_0 : \text{mset}(X) \to \text{mset}(Y), g_l : [\boxplus_{x \in l} A_x^{\otimes}] \to [\boxplus_{y \in g_0(l)} B_y^{\otimes}]),
\)
where $g_0 = \{ [x_0, \ldots, x_n] \mapsto [f_0(x_0), \ldots,
f_0(x_n)] \}$ and where $g_l$ is defined as shown on the right.

\begin{example}
  The lifting of the state boolean $b_s$ of Remark~\ref{rem:bool} to the
  parameter boolean $b_p$ is then a $\bbM$-map
  $\text{lb} : b_s\to F(b_p)$, where
  $F(b_p)$ is $(\text{mset}\{t\!t, f\!\!f\}, (\boxplus_{x\in
    l}I)_{l})$. The map $\text{lb}$ is defined as
  \( (\text{lb}_0, (f_x)_x) \)
  where $\text{lb}_0 : \{\emptyset\} \to \text{mset}\{t\!t, f\!\!f\}$
  sends $\emptyset$ to $[t\!t, f\!\!f]$, and where
  $\text{lb}_\emptyset :
  I\boxplus I\to I\boxplus I$ is simply defined as the identity.
  In the other direction, the $\bbM$-map $b_p\to b_s$ consists of the
  constant set-function on $\emptyset$ together with the
  injections $I\to I\boxplus I$ discussed
  in Section~\ref{sec:diag}.
\end{example}

\begin{remark}
  In Quipper dynamic lifting is implemented in the \texttt{Circ} monad
  which corresponds to the strong monad of $F$ in our model.  The
  branching side-effect corresponds to the \texttt{RW\_Read} constructor
  of the \texttt{Circ} monad.
\end{remark}

\subsection{Interpreting Typed Terms and Configurations}
\label{sec:kleisli}

In this section, we introduce an interpretation of Proto-Quipper-L
within the Kleisli category
$\bbMF$.
As it is customary, types are mapped to objects while typing
derivations are mapped to morphisms. When typed terms admit a unique
typing derivation this entails a unique denotation for typed terms. In
our situation, due to the promotion and dereliction rules typing
derivations are not necessarily unique: we therefore adjust the
statements of the lemmas and theorems accordingly. However, in the
case of values of basic types, thanks to Remark~\ref{rem:basic-type}
and the type safety properties, the denotation of closed terms of basic
types is independent from the choice of typing
derivation: this gives Corollary~\ref{cor:final}.

The interpretation $\denot{A}$ of a type $A$ is directly
built against the categorical structure:
$\denot{I} = (\{ \emptyset \}, (I))$,
$\denot{\text{bool}} = ( \{ \text{tt}, \text{ff} \}, (I,
I))$,
$\denot{\textbf{qubit}} = (\{ \emptyset \}, ([q]))$,
$\denot{A_a \multimap A_b} = \denot{A_a}
\multimap_{\bbMF} \denot{A_b}$ 
the internal hom
in the category $\bbMF$,
$\denot{A_a \otimes B_b} = \denot{A_a}
\otimes \denot{A_b}$,
$\denot{{!}A} = {!}\denot{A} = (p \circ
b)\denot{A}$. Finally, for quantum channels we follow
Rios\&Selinger's strategy by defining
$\denot{\text{QChan}(P, B)} =
p(\bbMF(\denot{P}, \denot{A}))$. In our situation, the set
$\bbMF(A,B)$ is isomorphic to $\bM(A,B)$
when $A$ and
$B$ are state objects: in this situation, QChan-types
indeed correspond to morphisms of the
category $\bM$,
i.e. quantum channels: this is used to interpret the box and unbox
operators. The quantum channel constant is just an
encapsulation over Definition~\ref{def:denot-qcalg}. 
Finally, a typed configuration ${!}\Delta \vdash (Q, m) : A$ is
interpreted as the composition of $Q$ (i.e. we first ``compute'' $Q$)
followed with the interpretation of $M$.

\begin{example}\label{ex:blah}
  The term exp of Eq.\eqref{eq:example-term} in
  Section~\ref{s:introduction} has for interpretation a morphism
  $(\{\emptyset\},(q))\to
  (\text{mset}\{(\emptyset,\emptyset)\},(q)_l)$ defined as
  \(
  (f_0, (f_\emptyset))
  \)
  where $f_0(\emptyset)=[(\emptyset,\emptyset),(\emptyset,\emptyset)]$
  and $f_\emptyset$ is defined as shown in Figure~\ref{fig:ex-den}
  (the dashed lines are meant to be vertical). The bottom
  box-node represents the measurement ($I\boxplus I$ being the result)
  and the upper one the test. The top
  result is a $\boxplus$-superposition
  of 2 copies of $q\otimes I$, as expected: these stand for the two
  ``classical'' possibilities
\end{example}

In general, soundness of categorical semantics states that the
categorical interpretation of the typing derivation is preserved over
the reduction. However, there can be multiple type derivations for 
each type judgement, in our type system, because of the reason explained 
above.  Therefore, in this paper, we show that for
a type judgement and a typing derivation, there exists a particular typing
judgement of the reduced type judgement which has the same
interpretation of the original typing derivation.

\begin{theorem}[Soundness]\label{th:sound}
  For any configurations $(Q_1, m_1)$ and $(Q_2, m_2)$ such that
  $(Q_1, m_1) \xrightarrow{} (Q_2, m_2)$, if $\vdash (Q_1, m_1) : A$,
  then for any typing derivation $\pi_1$ of $\vdash (Q_1, m_1) : A$,
  there exists a typing derivation $\pi_2$ of $\vdash (Q_2, m_2) : A$
  such that
  $\denot{\pi_1} = \denot{\pi_2}$.\qed
\end{theorem}

Finally, from the type safety properties, we can derive the following,
making it possible to define the interpretation of a closed term of
basic type.

\begin{corollary}\label{cor:final}
  All the typing derivations of a closed term of basic type share
  the same interpretation.\qed
\end{corollary}

\section{Conclusion}\label{s:discussion}

In this paper, we introduce the language Proto-Quipper-L which
formalizes several features of Quipper (dynamic lifting, 
higher-order function, circuit composition, and branching) while 
treating the qubits linearly using the type system.
On one hand we propose a type system and an operational
semantics which explains
the meaning of programs as a set of reduction rules.
On the other hand, we propose a concrete categorical model of the
language which is proven to be sound, meaning that the semantics is
preserved over the operational semantics.

On one side, the model is closely related to models of intuitionistic
linear logic. Diagrams are akin to proof nets: tensor nodes correspond
to multiplicative connectives while boxes correspond to additive
connectives.
On the other side, they can be considered as an extension of
diagrammatic languages for quantum
processes~\cite{staton2015algebraic}.

Our concrete semantics makes it possible to describe a monad,
following closely Quipper's operational semantics encoded in Haskell's
type system. With this semantics we are able to answer an open
question in the community: finding a categorical representation of
dynamic lifting for a circuit-description language.

\bibliography{biblio}

\newpage

\appendix

\section{Operational semantics}

\subsection{Reduction}
\label{app:definition-red}

The reduction rules for Proto-Quipper-L are defined as follows.

\subparagraph*{Reduction rules for classical computation}
The following rules always hold ($b$ is tt or ff)
\[
  \begin{aligned}
  (\epsilon(W), (\lambda x.M) V) &\xrightarrow{} (\epsilon(W), M[V / x])
  \\
  (\epsilon(W), \textbf{let}\ \langle x, y \rangle = \langle V, U \rangle\ \textbf{in}\ M) &\xrightarrow{} (\epsilon(W), M[V / x, U / y])
  \\
  (\epsilon(W), \textbf{if}\ \text{b}\ \textbf{then}\ M_{\text{tt}}\
  \textbf{else}\ M_{\text{ff}}) &\xrightarrow{} (\epsilon(W), M_{b})
\end{aligned}
\]

\subparagraph*{Reduction rules for circuit operations}
Provided that $\textbf{new}$ is an operator that creates free variables during
the computation, meaning that these free variables do not appear in both
classical and quantum contexts and that the term $\textbf{new}(P)$ is a
pattern of same shape as $P$ made out of these new variables, we have
\[
\inferrule
    {p = \textbf{new}(P) \\ W_p = \textbf{supp}(p)}
    {(\epsilon(\emptyset), \text{box}_P\,V) \xrightarrow{} (\epsilon(\emptyset), (p, \epsilon(W_p), V p))}
\hspace{1.5em}
\inferrule
    {\textbf{shape}(p) = \textbf{shape}(V) \\
     \sigma = \textbf{bind}(p, V)
     }
    {(\epsilon(\textbf{FV}(V)), (\text{unbox}(p,\ Q,\ u))V) \xrightarrow{} (\sigma(Q), \sigma(u))}
\]

\subparagraph*{Structural reduction rule for quantum channel constant}
Provided that $(Q, m) \xrightarrow{} (Q', m')$, we have 
$(\epsilon(\emptyset), (p, Q, m)) \xrightarrow{} (\epsilon(\emptyset), (p, Q', m'))$.

\subparagraph*{Structural reduction rules for empty quantum channel}
Provided that 
$(\epsilon(W_M),\ M) \xrightarrow{} (Q,\ m)$, that
$\textbf{all}(Q) \cap W_N = \emptyset$ and that $\textbf{all}(Q) \cap
W_V = \emptyset$, we have
\[
\begin{aligned}
(\epsilon(W_M \cup W_N),\ M N) &\xrightarrow{} (\textbf{extend}(Q,W_N),\ m N)
\\
(\epsilon(W \cup W_V),\ V M) &\xrightarrow{} (\textbf{extend}(Q,W_V),\
V m)
\\
(\epsilon(W_M \cup W_N),\ \langle M, N\rangle) &\xrightarrow{}
(\textbf{extend}(Q,W_N),\ \langle m, N\rangle)
\\
(\epsilon(W_M \cup W_V),\ \langle V, M\rangle) &\xrightarrow{}
(\textbf{extend}(Q,W_V),\ \langle V, m\rangle)
\\
(\epsilon(W_M \cup W_N),\ \textbf{if}\ M\ \textbf{then}\ M_a\ \textbf{else}\ M_b) ) &\xrightarrow{} (\textbf{extend}(Q,W_N),\ 
\textbf{if}\ m\ \textbf{then}\ M_a\ \textbf{else}\ M_b)
\\
(\epsilon(W_M),\ \textbf{let}\ \langle x, y \rangle =M \ \textbf{in}\ N)
    &\xrightarrow{} (\textbf{extend}(Q,W_N),\ 
    \textbf{let}\ \langle x, y \rangle = m \ \textbf{in}\ N)
  \end{aligned}
\]
We use syntactic sugar combining terms and branching terms, as in
$mM$. It corresponds to the term constructor applied to every leafs of
$m$, for instance: for $m = [[N_1, N_2], N_3]$, $[[N_1,N_2],N_3]M := [[N_1M,N_2M],N_3M]$.

\subparagraph*{Structural reduction rules for non-empty quantum channel}
Assume that $(Q_1,\ m_a) \xrightarrow{} (Q_3,\ m_c)$ and
$(Q_2,\ m_b) \xrightarrow{} (Q_4,\ m_d)$. Then
\[
  \begin{aligned}
    ((\text{meas}\ w\ Q_1\ Q_2),\ [m_a, m_b]) &\xrightarrow{}
    ((\text{meas}\ w\ Q_3\ Q_4),\ [m_c, m_d])
    \\
    ((\text{meas}\ w\ Q_1\ Q_2),\ [m_a, v]) &\xrightarrow{}
    ((\text{meas}\ w\ Q_3\ Q_2),\ [m_c, v])
    \\
    ((\text{meas}\ w\ Q_1\ Q_2),\ [v, m_b]) &\xrightarrow{}
    ((\text{meas}\ w\ Q_1\ Q_4),\ [v, m_d])
    \\
    (U(W)\ Q_1,\ m_a) &\xrightarrow{} (U(W) Q_3,\ m_c)
    \\
    (\text{init}\ b\ w\ Q_1,\ m_a) &\xrightarrow{} (\text{init}\ b\ w\ 
    Q_3,\ m_c)
    \\
    (\text{free}\ w\ Q_1,\ m_a) &\xrightarrow{} (\text{free}\ w\ 
    Q_3,\ m_c)
  \end{aligned}
\]

\subsection{Derivation of the example of Example~\ref{sec:example-red}}
\label{app:example-red}

Let us explain how the tree expands as the computation
progresses for example~\ref{sec:example-red}. First, we show that
$((\epsilon\{ \}, \text{init}(\text{tt})) \xrightarrow{}
((\text{init}\ \text{true}\ x\ \epsilon\{ x \}, x)$ as follows.
\[
\inferrule*
  {\textbf{shape}(*) = \textbf{shape}(*) \\
  \sigma = \textbf{bind}(*,*)}
  {
  \begin{tikzpicture}[baseline=-0.5ex,thick]
  \draw (0,0) node[] (qs) {};
  \draw (1.4,0) node[rectangle, draw] (qt) {$\text{init}(\text{tt})$};
  %
  \path[every node/.style={font=\sffamily\small, inner sep=1pt}]
  (qs) edge [-, sloped, pos=0.6, above] node {} (qt);
  %
  \begin{pgfonlayer}{background} 
  \node[surround] (background) [fit = (qs)] {};
  \end{pgfonlayer}
  \end{tikzpicture}
  \quad \xrightarrow{} \quad
  \begin{tikzpicture}[baseline=-0.5ex,thick]
  %
  \draw (0,0) node[] (qs) {$*$};
  \draw (1.3,0) node[operator] (qo) {$\text{init}\ \text{true}\ x$};
  \draw (3,0) node[rectangle, draw] (qt) {$x$};
  %
  \path[every node/.style={font=\sffamily\small, inner sep=1pt}]
  (qs) edge [-, sloped, above] node {} (qo)
  (qo) edge [-, sloped, pos=0.6, above] node {$x$} (qt);
  %
  \begin{pgfonlayer}{background} 
  \node[surround] (background) [fit = (qs) (qo)] {};
  \end{pgfonlayer}
  \end{tikzpicture} \sim
  \begin{tikzpicture}[baseline=-0.5ex,thick]
  %
  \draw (0,0) node[] (qs) {$*$};
  \draw (1.3,0) node[operator] (qo) {$\text{init}\ \text{true}\ v_d$};
  \draw (3.2,0) node[rectangle, draw] (qt) {$v_d$};
  %
  \path[every node/.style={font=\sffamily\small, inner sep=1pt}]
  (qs) edge [-, sloped, above] node {} (qo)
  (qo) edge [-, sloped, pos=0.6, above] node {$v_d$} (qt);
  %
  \begin{pgfonlayer}{background} 
  \node[surround] (background) [fit = (qs) (qo)] {};
  \end{pgfonlayer}
  \end{tikzpicture}
  }
\]
where we let
\[
\text{init}(\text{tt})\quad {=}\quad
\text{unbox}
\left(
*,\ 
\begin{tikzpicture}[baseline=-0.5ex,thick]
  \tikzstyle{operator} = [draw,fill=white,minimum size=1.5em] 
  %
  \draw (0,0) node[] (qs) {$*$};
  \draw (1.3,0) node[operator] (qi) {$\text{init}\ \text{true}\ x$};
  \draw (2.7,0) node[] (qt) {};
  %
  \path[every node/.style={font=\sffamily\small, inner sep=1pt}]
  (qs) edge [-, sloped] node {} (qi)
  (qi) edge [-, sloped, pos=0.7, above] node {$x$} (qt);
  %
  \begin{pgfonlayer}{background} 
  \node[surround] (background) [fit = (qs) (qi)] {};
  \end{pgfonlayer}
\end{tikzpicture},\ x
\right) (*).
\]
Then we can show the following reduction:
\[
\inferrule*
  {
  \inferrule*
    {
    \begin{tikzpicture}[baseline=-0.5ex,thick]
    \draw (0,0) node[] (qs) {};
    \draw (1.4,0) node[rectangle, draw] (qt) {$\text{init}(\text{tt})$};
    %
    \path[every node/.style={font=\sffamily\small, inner sep=1pt}]
    (qs) edge [-, sloped, pos=0.6, above] node {} (qt);
    %
    \begin{pgfonlayer}{background} 
    \node[surround] (background) [fit = (qs)] {};
    \end{pgfonlayer}
    \end{tikzpicture} \xrightarrow{}
    \begin{tikzpicture}[baseline=-0.5ex,thick]
    %
    \draw (0,0) node[] (qs) {$*$};
    \draw (1.3,0) node[operator] (qo) {$\text{init}\ \text{true}\ v_d$};
    \draw (3.2,0) node[rectangle, draw] (qt) {$v_d$};
    %
    \path[every node/.style={font=\sffamily\small, inner sep=1pt}]
    (qs) edge [-, sloped, above] node {} (qo)
    (qo) edge [-, sloped, pos=0.6, above] node {$v_d$} (qt);
    %
    \begin{pgfonlayer}{background} 
    \node[surround] (background) [fit = (qs) (qo)] {};
    \end{pgfonlayer}
    \end{tikzpicture} \\
    \textbf{all}(\text{init}\ \text{true}\ v_d\ \epsilon\{ v_d \}) \cap \{ v_c \} = \emptyset
    }
    {
    \begin{tikzpicture}[baseline=-0.5ex,thick]
    %
    \draw (0,0) node[] (qs) {};
    \draw (2.2,0) node[rectangle, draw] (qt) {$\langle \text{init}(\text{tt}), \text{free}(v_c) \rangle$};
    %
    \path[every node/.style={font=\sffamily\small, inner sep=1pt}]
    (qs) edge [-, sloped, pos=0.6, above] node {$v_c$} (qt);
    %
    \begin{pgfonlayer}{background} 
    \node[surround] (background) [fit = (qs)] {};
    \end{pgfonlayer}
    \end{tikzpicture}
    \quad \xrightarrow{} \quad
    \begin{tikzpicture}[baseline=-0.5ex,thick]
    %
    \draw (0,0) node[] (qs) {};
    \draw (1.7,0) node[operator] (qo) {$\text{init}\ \text{true}\ v_d$};
    \draw (5,0) node[rectangle, draw] (qt) {$\langle v_d, \text{free}(v_c) \rangle$};
    %
    \path[every node/.style={font=\sffamily\small, inner sep=1pt}]
    (qs) edge [-, sloped, above] node {$v_c$} (qo)
    (qo) edge [-, sloped, pos=0.6, above] node {$v_c, v_d$} (qt);
    %
    \begin{pgfonlayer}{background} 
    \node[surround] (background) [fit = (qs) (qo)] {};
    \end{pgfonlayer}
    \end{tikzpicture}
    }
  }
  {
  \begin{tikzpicture}[baseline=-0.5ex,thick]
  %
  \draw (0,0) node[] (qs) {};
  \draw (1.2,0) node[operator] (qm) {$\text{meas}\ v_c$};
  \draw (3.6,0.7) node[rectangle, draw] (qt1) {$\langle \text{init}(\text{tt}), \text{free}(v_c) \rangle$};
  \draw (2.7,-0.7) node[rectangle, draw] (qt2) {$\langle v_c, * \rangle$};
  %
  \path[every node/.style={font=\sffamily\small, inner sep=1pt}]
  (qs) edge [-, sloped, above] node {$v_c$} (qm)
  (qm) [bend left=30] edge [-, sloped, pos=0.7, above] node {$v_c$} (qt1.west)
  (qm) [bend right=30] edge [-, sloped, pos=0.7, above] node {$v_c$} (qt2.west);
  %
  \begin{pgfonlayer}{background} 
  \node[surround] (background) [fit = (qs) (qm)] {};
  \end{pgfonlayer}
  \end{tikzpicture}
  \quad \xrightarrow{} \quad
  \begin{tikzpicture}[baseline=-0.5ex,thick]
  %
  \draw (0,0.-0.5) node[] (qs) {};
  \draw (1.2,-0.5) node[operator] (qm) {$\text{meas}\ v_c$};
  \draw (1.7,0.5) node[operator] (qi) {$\text{init}\ \text{true}\ v_d$};
  \draw (4.9,0.5) node[rectangle, draw] (qt1) {$\langle v_d, \text{free}(v_c) \rangle$};
  \draw (4.3,-0.5) node[rectangle, draw] (qt2) {$\langle v_c, * \rangle$};
  %
  \path[every node/.style={font=\sffamily\small, inner sep=1pt}]
  (qs) edge [-, sloped, above] node {$v_c$} (qm)
  (qm)  edge [-, sloped, bend left=30, pos=0.2, above] node {$v_c$} (qi.south)
  (qi) edge [-, sloped, pos=0.55, above] node {$v_c, v_d$} (qt1)
  (qm) edge [-, sloped, pos=0.9, above] node {$v_c$} (qt2);
  %
  \begin{pgfonlayer}{background} 
  \node[surround] (background) [fit = (qs) (qm) (qi)] {};
  \end{pgfonlayer}
  \end{tikzpicture}
  }
\]

Next, we show the last reduction step of the example.
\[
\inferrule*
  {\textbf{shape}(x) = \textbf{shape}(v_c) \\ \sigma = \textbf{bind}(x, v_c)}
  {
  \begin{tikzpicture}[baseline=-0.5ex,thick]
  \draw (0,0) node[] (qs) {};
  \draw (1.6,0) node[rectangle, draw] (qt) {$\text{free}(v_c)$};
  %
  \path[every node/.style={font=\sffamily\small, inner sep=1pt}]
  (qs) edge [-, sloped, pos=0.6, above] node {$v_c$} (qt);
  %
  \begin{pgfonlayer}{background} 
  \node[surround] (background) [fit = (qs)] {};
  \end{pgfonlayer}
  \end{tikzpicture}
  \quad \xrightarrow{} \quad
  \begin{tikzpicture}[baseline=-0.5ex,thick]
  %
  \draw (0,0) node[] (qs) {};
  \draw (1.4,0) node[operator] (qo) {$\text{free}\ v_c$};
  \draw (3,0) node[rectangle, draw] (qt) {$*$};
  %
  \path[every node/.style={font=\sffamily\small, inner sep=1pt}]
  (qs) edge [-, sloped, above] node {$v_c$} (qo)
  (qo) edge [-, sloped, pos=0.6, above] node {} (qt);
  %
  \begin{pgfonlayer}{background} 
  \node[surround] (background) [fit = (qs) (qo)] {};
  \end{pgfonlayer}
  \end{tikzpicture}
  }
\]
Recall that 
\[
\text{free}\quad {=}\quad
\text{unbox}
\left(
x,\ 
\begin{tikzpicture}[baseline=-0.5ex,thick]
  \tikzstyle{operator} = [draw,fill=white,minimum size=1.5em] 
  %
  \draw (0,0) node[] (qs) {};
  \draw (1.1,0) node[operator] (qf) {$\text{free}\ x$};
  \draw (2.2,0) node[] (qt) {$*$};
  %
  \path[every node/.style={font=\sffamily\small, inner sep=1pt}]
  (qs) edge [-, sloped, above] node {$x$} (qf)
  (qf) edge [-, sloped] node {} (qt);
  %
  \begin{pgfonlayer}{background} 
  \node[surround] (background) [fit = (qs) (qf)] {};
  \end{pgfonlayer}
\end{tikzpicture},\ *
\right).
\]
Then we can show the following reduction:

\scalebox{.85}{$
\inferrule*
  {
  \inferrule*
    {
    \inferrule*
      {
      \begin{tikzpicture}[baseline=-0.5ex,thick]
      \draw (0,0) node[] (qs) {};
      \draw (1.6,0) node[rectangle, draw] (qt) {$\text{free}(v_c)$};
      %
      \path[every node/.style={font=\sffamily\small, inner sep=1pt}]
      (qs) edge [-, sloped, pos=0.6, above] node {$v_c$} (qt);
      %
      \begin{pgfonlayer}{background} 
      \node[surround] (background) [fit = (qs)] {};
      \end{pgfonlayer}
      \end{tikzpicture}
      \xrightarrow{}
      \begin{tikzpicture}[baseline=-0.5ex,thick]
      %
      \draw (0,0) node[] (qs) {};
      \draw (1.4,0) node[operator] (qo) {$\text{free}\ v_c$};
      \draw (3,0) node[rectangle, draw] (qt) {$*$};
      %
      \path[every node/.style={font=\sffamily\small, inner sep=1pt}]
      (qs) edge [-, sloped, above] node {$v_c$} (qo)
      (qo) edge [-, sloped, pos=0.6, above] node {} (qt);
      %
      \begin{pgfonlayer}{background} 
      \node[surround] (background) [fit = (qs) (qo)] {};
      \end{pgfonlayer}
      \end{tikzpicture}
      \\
      \textbf{all}(\text{free}\ v_c\ \epsilon\{ \}) \cap \{ v_d \} = \emptyset
      }
      {
      \begin{tikzpicture}[baseline=-0.5ex,thick]
      %
      \draw (0,0) node[] (qs) {};
      \draw (2.4,0) node[rectangle, draw] (qt) {$\langle v_d, \text{free}(v_c) \rangle$};
      %
      \path[every node/.style={font=\sffamily\small, inner sep=1pt}]
      (qs) edge [-, sloped, pos=0.6, above] node {$v_c, v_d$} (qt);
      %
      \begin{pgfonlayer}{background} 
      \node[surround] (background) [fit = (qs)] {};
      \end{pgfonlayer}
      \end{tikzpicture}
      \quad \xrightarrow{} \quad
      \begin{tikzpicture}[baseline=-0.5ex,thick]
      %
      \draw (0,0) node[] (qs) {};
      \draw (1.8,0) node[operator] (qo) {$\text{free}\ v_c$};
      \draw (3.8,0) node[rectangle, draw] (qt) {$\langle v_d, * \rangle$};
      %
      \path[every node/.style={font=\sffamily\small, inner sep=1pt}]
      (qs) edge [-, sloped, above] node {$v_c, v_d$} (qo)
      (qo) edge [-, sloped, pos=0.6, above] node {$v_d$} (qt);
      %
      \begin{pgfonlayer}{background} 
      \node[surround] (background) [fit = (qs) (qo)] {};
      \end{pgfonlayer}
      \end{tikzpicture}
      }
    }
    {
    \begin{tikzpicture}[baseline=-0.5ex,thick]
    %
    \draw (0,0) node[] (qs) {};
    \draw (1.7,0) node[operator] (qo) {$\text{init}\ \text{true}\ v_d$};
    \draw (5,0) node[rectangle, draw] (qt) {$\langle v_d, \text{free}(v_c) \rangle$};
    %
    \path[every node/.style={font=\sffamily\small, inner sep=1pt}]
    (qs) edge [-, sloped, above] node {$v_c$} (qo)
    (qo) edge [-, sloped, pos=0.6, above] node {$v_c, v_d$} (qt);
    %
    \begin{pgfonlayer}{background} 
    \node[surround] (background) [fit = (qs) (qo)] {};
    \end{pgfonlayer}
    \end{tikzpicture}
    \quad \xrightarrow{} \quad
    \begin{tikzpicture}[baseline=-0.5ex,thick]
    %
    \draw (0,0) node[] (qs) {};
    \draw (1.8,0) node[operator] (qi) {$\text{init}\ \text{true}\ v_d$};
    \draw (4.6,0) node[operator] (qf) {$\text{free}\ v_c$};
    \draw (6.6,0) node[rectangle, draw] (qt) {$\langle v_d, * \rangle$};
    %
    \path[every node/.style={font=\sffamily\small, inner sep=1pt}]
    (qs) edge [-, sloped, above] node {$v_c$} (qi)
    (qi) edge [-, sloped, above] node {$v_c, v_d$} (qf)
    (qf) edge [-, sloped, pos=0.6, above] node {$v_d$} (qt);
    %
    \begin{pgfonlayer}{background} 
    \node[surround] (background) [fit = (qs) (qi) (qf)] {};
    \end{pgfonlayer}
    \end{tikzpicture}
    }
  }
  {
  \begin{tikzpicture}[baseline=-0.5ex,thick]
  %
  \draw (0,0.-0.5) node[] (qs) {};
  \draw (1.2,-0.5) node[operator] (qm) {$\text{meas}\ v_c$};
  \draw (1.7,0.5) node[operator] (qi) {$\text{init}\ \text{true}\ v_d$};
  \draw (4.9,0.5) node[rectangle, draw] (qt1) {$\langle v_d, \text{free}(v_c) \rangle$};
  \draw (4.4,-0.5) node[rectangle, draw] (qt2) {$\langle v_c, * \rangle$};
  %
  \path[every node/.style={font=\sffamily\small, inner sep=1pt}]
  (qs) edge [-, sloped, above] node {$v_c$} (qm)
  (qm)  edge [-, sloped, bend left=30, pos=0.2, above] node {$v_c$} (qi.south)
  (qi) edge [-, sloped, pos=0.55, above] node {$v_c, v_d$} (qt1)
  (qm) edge [-, sloped, pos=0.9, above] node {$v_c$} (qt2);
  %
  \begin{pgfonlayer}{background} 
  \node[surround] (background) [fit = (qs) (qm) (qi)] {};
  \end{pgfonlayer}
  \end{tikzpicture}
  \quad \xrightarrow{} \quad
  \begin{tikzpicture}[baseline=-0.5ex,thick]
  %
  \draw (0,-0.5) node[] (qs) {};
  \draw (1.2,-0.5) node[operator] (qm) {$\text{meas}\ v_c$};
  \draw (1.7,0.5) node[operator] (qi) {$\text{init}\ \text{true}\ v_d$};
  \draw (4.4,0.5) node[operator] (qf) {$\text{free}\ v_c$};
  \draw (6.4,0.5) node[rectangle, draw] (qt1) {$\langle v_d, * \rangle$};
  \draw (6.4,-0.5) node[rectangle, draw] (qt2) {$\langle v_c, * \rangle$};
  %
  \path[every node/.style={font=\sffamily\small, inner sep=1pt}]
  (qs) edge [-, sloped, above] node {$v_c$} (qm)
  (qm)  edge [-, sloped, bend left=30, pos=0.2, above] node {$v_c$} (qi.south)
  (qi) edge [-, sloped, above] node {$v_c, v_d$} (qf)
  (qf) edge [-, sloped, pos=0.6, above] node {$v_d$} (qt1)
  (qm) edge [-, sloped, pos=0.95, above] node {$v_c$} (qt2);
  %
  \begin{pgfonlayer}{background} 
  \node[surround] (background) [fit = (qs) (qm) (qi) (qf)] {};
  \end{pgfonlayer}
  \end{tikzpicture}
  }
$}

\section{Categorical semantics}

\subsection{Equivalence of diagrams}

\begin{figure}[tb]
\scalebox{0.9}{
\includegraphics[width=0.45\textwidth]{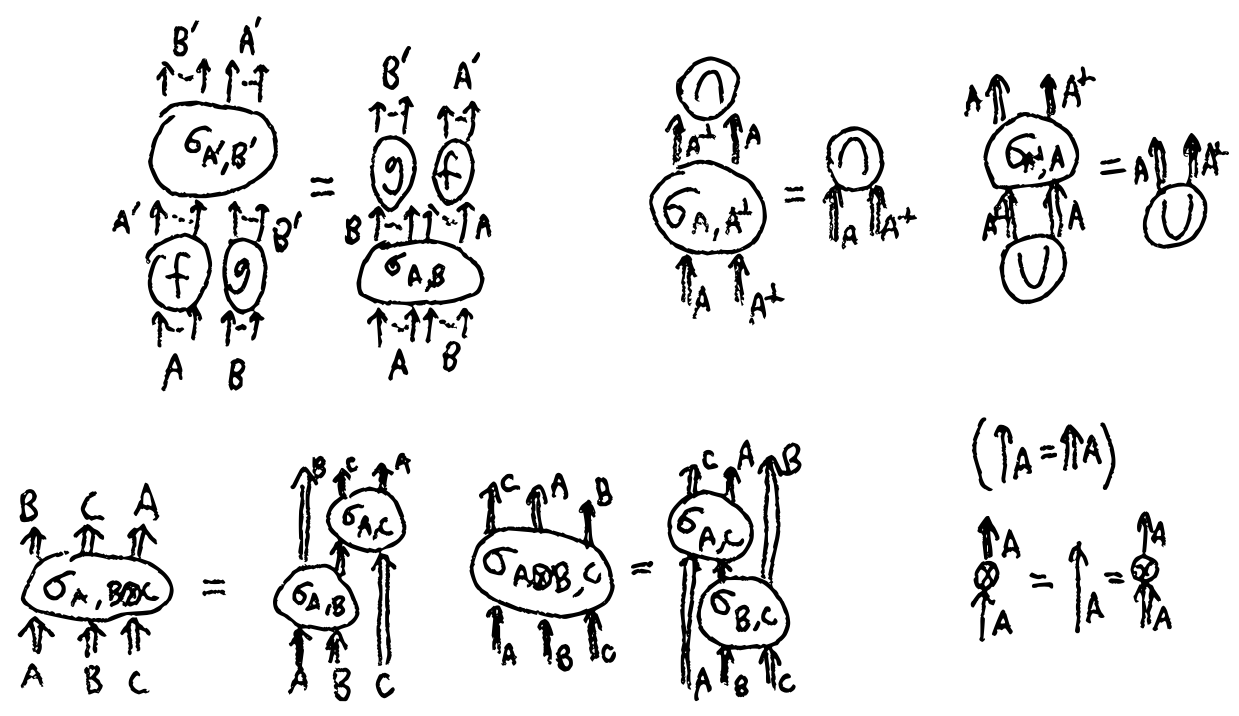}
\includegraphics[width=0.45\textwidth]{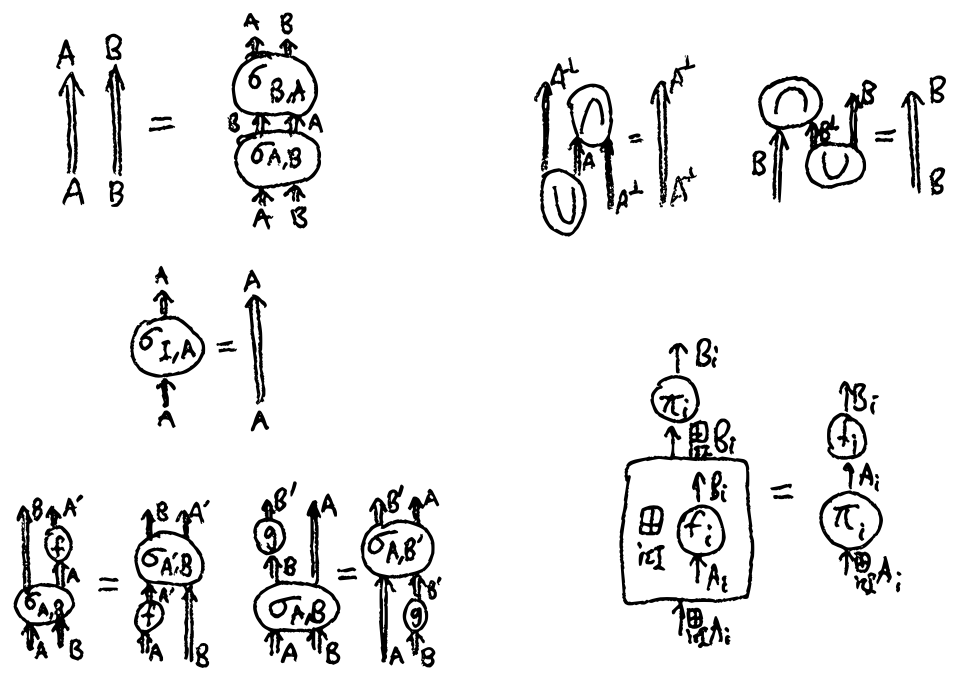}
}

\scalebox{0.9}{
\includegraphics[width=0.2\textwidth]{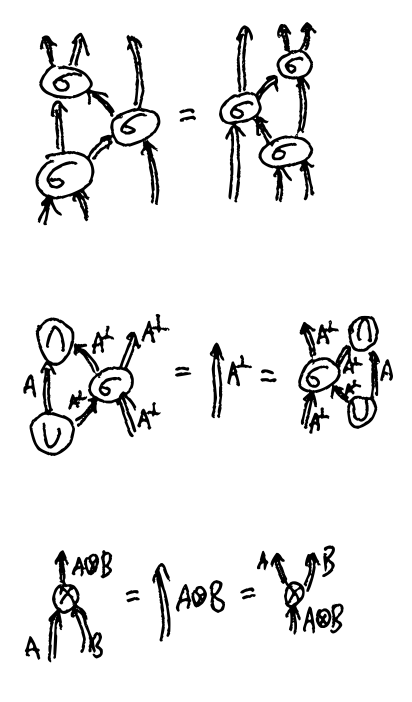}
\includegraphics[width=0.2\textwidth]{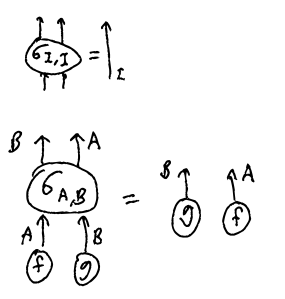}
\includegraphics[width=0.1\textwidth]{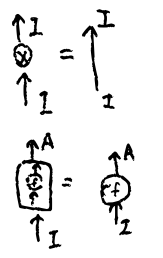}
\includegraphics[width=0.15\textwidth]{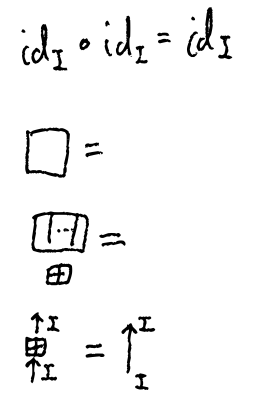}
\includegraphics[width=0.2\textwidth]{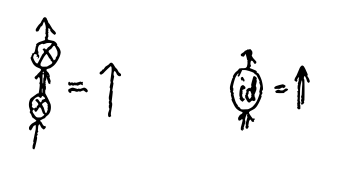}
}

\scalebox{0.9}{
\includegraphics[width=0.4\textwidth]{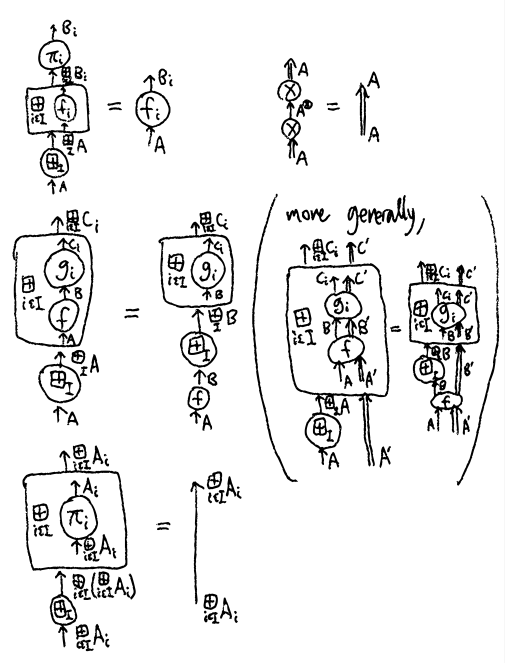}
\includegraphics[width=0.4\textwidth]{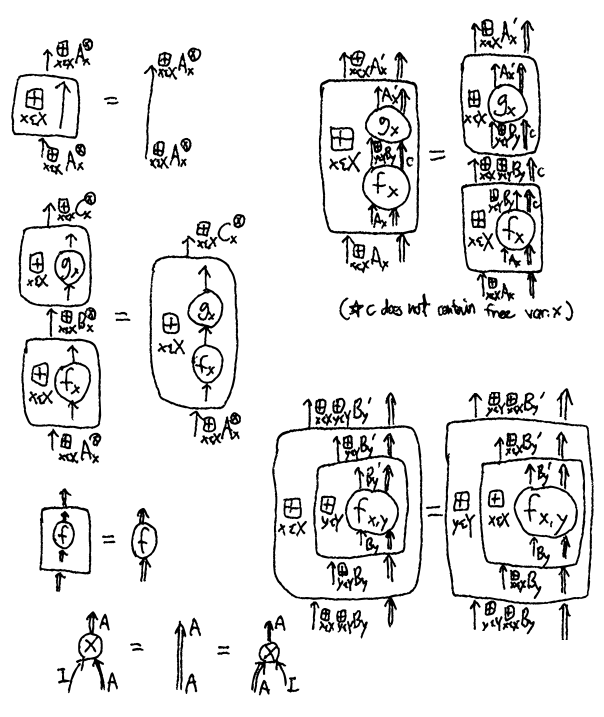}
\includegraphics[width=0.1\textwidth]{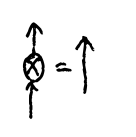}
}

\scalebox{0.9}{
\includegraphics[width=0.4\textwidth]{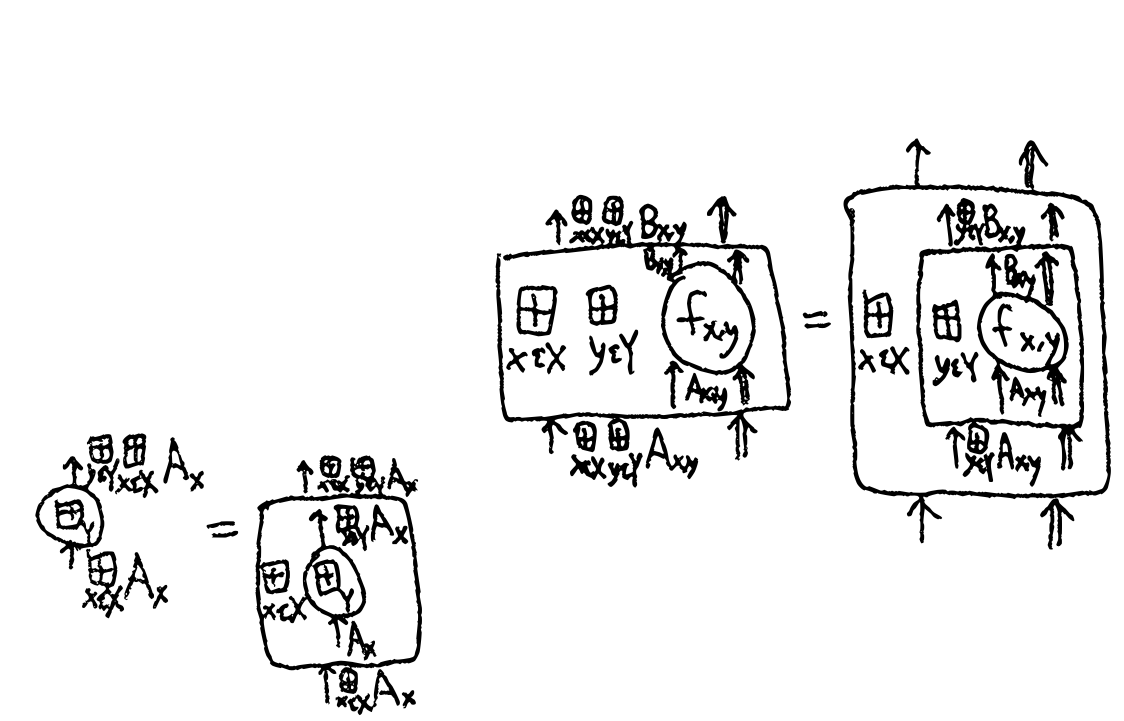}
\includegraphics[width=0.4\textwidth]{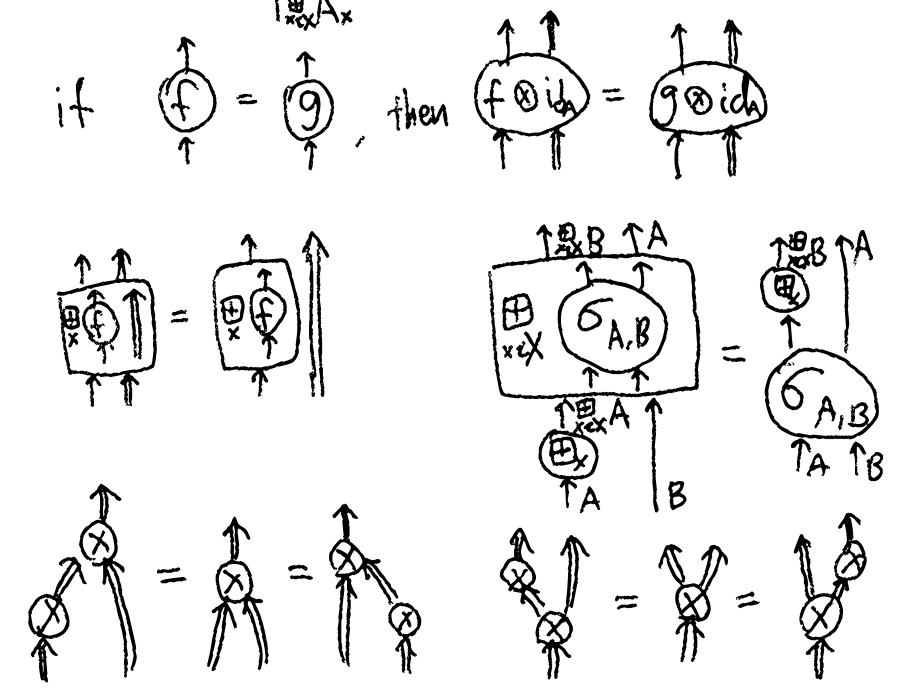}
\includegraphics[width=0.1\textwidth]{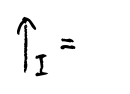}
}

\caption{Equivalence relation of diagrams}
\label{fig:equiv_diagram}
\end{figure}

Complete list of the equivalence rules that are used to construct the 
categorical model is shown in Figure~\ref{fig:equiv_diagram}.

\subsection{Interpretation of type system}

For a typing context
$\Gamma = x_1 : A_1, \ldots, x_k : A_k$, assuming the variables are
ordered by some linear order, 
$\denot{\Gamma} = \denot{A_1} \otimes \ldots \otimes \denot{A_k}$.
Next, the branching typing context is interpreted as the coproduct of the objects assigned to the smaller branching typing contexts, namely:
$\denot{\gamma_1, \gamma_2} = \denot{\gamma_1} + \denot{\gamma_2}$.
Lastly, we interpret the typing derivation as a morphism in
$\bbMF$. 

\subsubsection{Quantum channel types, Box and Unbox}

As in \cite{rios2017categorical}, we interpret the quantum
channel types $\textbf{QChan}(A, B)$ as an object
$p(\bbMF(A, B)) = (\bbMF(A, B), (I))$ in
$\bbMF$ and $\bbM$.  Note that the object
is a parameter object as in \cite{rios2017categorical}, which means
that the object has the form of $(X, (I)_X)$ for some $X$.
When we define the quantum channel types $\textbf{QChan}(A, B)$ as a
parameter object, box and unbox can be interpreted based on an
isomorphism between the set
$b(A \multimap_{\bbM} B)$ and
$\bbM(A, B)$.  In specific, we can define an
isomorphism as in Table~\ref{table:def_isomorphism}.

\begin{table}[tb]
  \begin{tabular}{| p{0.5\textwidth} | p{0.4\textwidth} |}
  \hline
  $\text{iso}_{\xrightarrow{}} : b(A \multimap_{\bbM} B) \to \bbM(A, B)$:
  &
  $\text{iso}_{\xleftarrow{}} : \bbM(A, B) \to b(A \multimap_{\bbM} B)$:
  \\ \hline ~&~\\[-2ex]
  Given $(f, d_f)$, which is $\left(f, \raisebox{-0.5\height}{\includegraphics[width=0.1\textwidth]{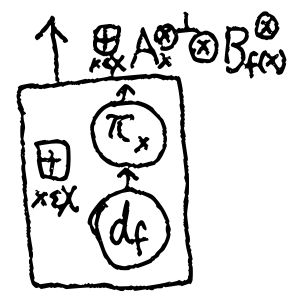}}\right)$, let 
  &
  Given $(f_0, (f_x)_X)$, let 
  \\
  \[
  f_0 = f\ \text{and}\ f_x = \raisebox{-0.5\height}{\includegraphics[width=0.1\textwidth]{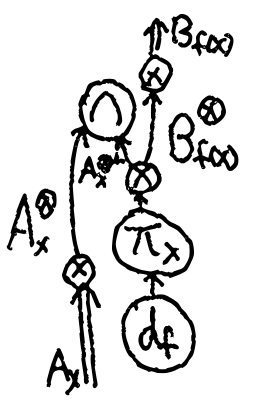}}.
  \]
  &
  \[
  f = f_0\ \text{and}\ d_f = \raisebox{-0.5\height}{\includegraphics[width=0.1\textwidth]{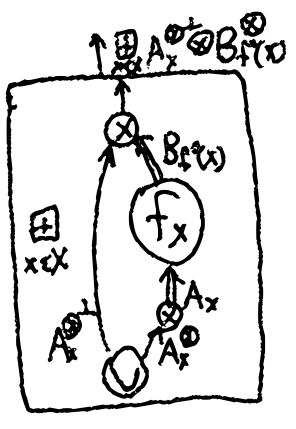}}
  \]
  \\ \hline
  \end{tabular}
  \caption{Definition of isomorphism between $b(A \multimap_{\bbM} B)$ and $\bbM(A, B)$}
  \label{table:def_isomorphism}
  \vspace{-1em}
\end{table}

Given the isomorphism, we can define the morphisms for box and unbox
as morphisms in $\bbM$ as follows:
\[
\begin{aligned}
\text{unbox} & = p(\bbM(A, F(B)) \xrightarrow{p(\text{iso}_{\xrightarrow{}})} (p \circ b) (A \multimap_{\bbM} F(B)) \xrightarrow{\epsilon(A \multimap_{\bbM} F(B))} (A \multimap_{\bbM} F(B)) \\
\text{box} & = (p \circ b)(A \multimap_{\bbM} F(B))
\xrightarrow{p(\text{iso}_{\xleftarrow{}})} p(\bbM(A, F(B)))\\&\qquad
\qquad\qquad\xrightarrow{p(\eta(\bbM(A, F(B))))} (p \circ b \circ p)(\bbM(A, F(B))).
\end{aligned}
\]
where $\epsilon$ refers to the counit from the comonad $!$.

\subsubsection{Quantum channel constants}

We define a natural transformations called $\text{bif}$
and $\text{merge}$ for the measurement as in Table~\ref{table:def_bif_and_merge}.

A quantum channel $Q$ is interpreted as a morphism
$\denot{\textbf{in}(Q)} \xrightarrow{}_{\bbMF}
\denot{\textbf{out}(Q)}$, where
\[
\begin{aligned}
\denot{\textbf{in}(Q)} & = (\{ \emptyset \}, ([q]^{\otimes |\textbf{in}(Q)|})) \\
\denot{\textbf{out}(Q)} & = 
\begin{cases}
  (\{ \emptyset \}, ([q]^{\otimes |V|})) & \text{if $\textbf{out}(Q)$ is a set $V$} \\
  \denot{o_1} + \denot{o_2} & \text{if $\textbf{out}(Q) = [o_1, o_2]$}
  \end{cases}
\end{aligned}
\]

The interpretation of quantum channel is defined inductively as in
Table~\ref{table:interpretation_qc_constant} where $\mu$ represents
the multiplication of the monad $F$. In addition, the elementary nodes in 
Table~\ref{table:interpretation_qc_constant}--$(U(V_1))$, $(\text{free}\ v)$, $(\mid b \rangle,\ v)$ and $(\langle b \mid,\ v)$--refers to the unitary
gate node $\rbox{U}$ (which is either $1$ or $2$-qubits gate) applied to wires $V_1$, $\rbox{\text{tr}}$ node applied to wire $v$, and $\rbox{\langle b|}$ and $\rbox{|b\rangle}$ nodes applied to wire $v$, respectively.

\begin{table}[!tb]
  \begin{tabular}{| p{0.425\textwidth} | p{0.425\textwidth} |}
  \hline
  $\text{bif}(A) : A \xrightarrow{}_{\bbMF} A + A$
  &
  $\text{merge}(A, B) : F(A) + F(B) \xrightarrow{}_{\bbMF}
  A + B$
  \\
  \hline
  For an object $A = (X, (A_x))$, we let 
  \[
  \begin{aligned}
  & \text{bif}(X, (A_x)) = \\
  & \left(\begin{array}{l}\{ x \mapsto [(0, x), (1, x)] \},\\\qquad (f_x : A_x \to A_x^{\otimes} \boxplus A_x^{\otimes})\end{array}\right)
  \end{aligned}
  \]
  where $f_x =
    \raisebox{-1cm}{\includegraphics[width=1.5cm]{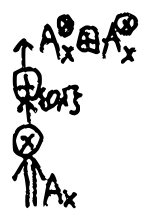}}$.
  &
  For objects $A, B$, we let
  \[
  \begin{aligned}
  & \text{merge}(A, B) = (\{ \\
  & (0, [x_1, \ldots, x_k]) \mapsto [(0, x_1), \ldots, (0, x_k)], \\
  & (1, [y_1, \ldots, y_n]) \mapsto [(1, y_1), \ldots, (1, y_n)] \}, \\
  & (\textbf{id}_{[\boxplus_{x \in l} A_x^{\otimes}]})_{l : \textbf{mset}(X)}\\&\quad {+}{+} (\textbf{id}_{[\boxplus_{y \in l} B_y^{\otimes}]})_{l :
    \textbf{mset}(Y)})
  \end{aligned}
  \]
  \\
  \hline
  It satisfies the following commute diagram for naturality:
  &
  It satisfies the following commute diagram for naturality:
  \\
  \begin{tikzcd}
    A \arrow[d, "f"] \arrow[r, "\text{bif}(A)"] & F(A + A) \arrow[d, "F(f + f)"] \\
    B \arrow[r, "\text{bif}(B)"] & F(B + B)
  \end{tikzcd}
  &
  \begin{tikzcd}
    F(A) + F(B) \arrow[d, "F(f) + F(g)"] \arrow[r, "\text{merge}(A \text{,} B)"] & F(A + B) \arrow[d, "F(f + g)"] \\
    F(A') + F(B') \arrow[r, "\text{merge}(A' \text{,} B')"] & F(A' + B')
  \end{tikzcd}
  \\
  \hline
  \end{tabular}
  \caption{Definition of $\text{bif}$ and $\text{merge}$}
  \label{table:def_bif_and_merge}
  \vspace{-1em}
\end{table}

\begin{table}[!tb]
  \begin{tabular}{| p{0.5\textwidth} | p{0.35\textwidth} |}
  \hline
  \begin{minipage}{0.49\textwidth}
  \[
  \begin{aligned}
  \denot{\epsilon(V)} & = \eta(\{ \emptyset \}, ([q]^{\otimes |V|})) \\
  \denot{U(V_1)\ Q} & = \denot{Q} \circ \denot{U(V_1)}^{0} \\
  \denot{\text{free}\ v\ Q} & = \denot{Q} \circ \denot{\text{free}(V)}^{0} \\
  \denot{\text{init}\ b\ v\ Q} & = \denot{Q} \circ \denot{\text{init}(b, v)}^{0} \\
  \denot{\text{meas}\ v\ Q_1\ Q_2} & = \\
  &\hspace{-2cm}\begin{tikzcd}
  \denot{\textbf{in}(\text{meas}\ v\ Q_1\ Q_2)} \arrow[d, "\text{bif}; F\left(\begin{array}{l}\denot{Q_1} \circ \denot{\text{meas}(v \text{,} 0)}^{0}\\ + \denot{Q_2} \circ \denot{\text{meas}(v \text{,} 1)}^{0}\end{array}\right)"] \\
  F(F\denot{\textbf{out}(Q_1)} + F\denot{\textbf{out}(Q_2)}) \arrow[d, "F(\text{merge}); \mu"] \\
  F(\denot{\textbf{out}(Q_1)} + \denot{\textbf{out}(Q_2)})
  \end{tikzcd}
  \end{aligned}
  \]
\end{minipage}
                      &
  \begin{minipage}{0.34\textwidth}where\\
    \includegraphics[width=1.8in]{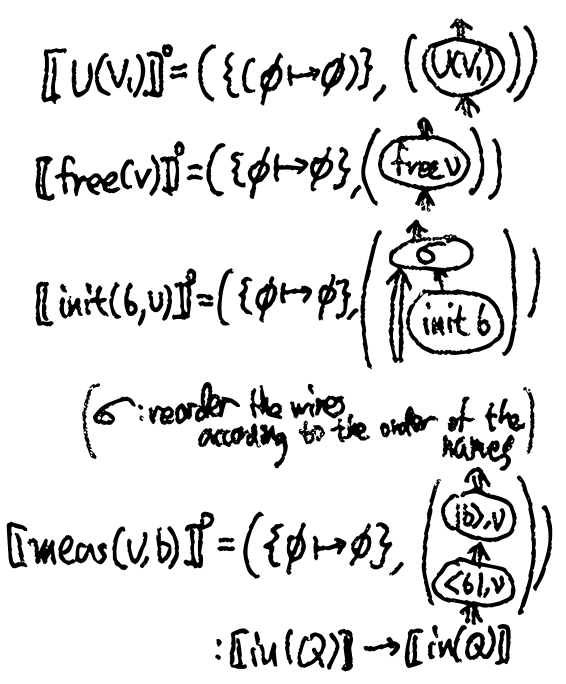}
  \end{minipage}
    \\ \hline
  \end{tabular}
  \caption{Interpretation of quantum channel constants}
  \label{table:interpretation_qc_constant}
  \vspace{-1em}
\end{table}

\end{document}